\documentclass[]{emulateapj}
\pdfoutput=1

\newcommand{\unit}[1]{\ensuremath{\, \mathrm{#1}}}

\usepackage{graphicx}
\usepackage[backref,breaklinks,colorlinks,citecolor=blue]{hyperref}
\usepackage[all]{hypcap}

\usepackage{color}

\begin{document}

\title{Recovery of Large Angular Scale CMB Polarization for Instruments Employing Variable-delay Polarization Modulators}
\author{N.J. Miller\altaffilmark{1,2,*}, D.T. Chuss\altaffilmark{3}, T.A. Marriage\altaffilmark{1}, E.J. Wollack\altaffilmark{2}, J.W. Appel\altaffilmark{1}, 
C.L. Bennett\altaffilmark{1}, J. Eimer\altaffilmark{1}, T. Essinger-Hileman\altaffilmark{1}, D.J. Fixsen\altaffilmark{2}, K. Harrington\altaffilmark{1}, 
S.H. Moseley\altaffilmark{2}, K. Rostem\altaffilmark{1,2}, E.R. Switzer\altaffilmark{2}, D.J. Watts\altaffilmark{1}}
\altaffiltext{1}{Department of Physics and Astronomy, Johns Hopkins University, 3400 N. Charles St., Baltimore, MD 21218}
\altaffiltext{2}{Observational Cosmology Laboratory, Code 665, NASA Goddard Space Flight Center, Greenbelt, MD 20771}
\altaffiltext{3}{Department of Physics, Villanova University, 800 E Lancaster, Villanova, PA 19085}
\altaffiltext{*}{\href{mailto:Nathan.J.Miller@nasa.gov}{Nathan.J.Miller@nasa.gov}}

\begin{abstract}
Variable-delay Polarization Modulators (VPMs) are currently being implemented in experiments designed to measure the 
polarization of the cosmic microwave background on large angular scales because of their capability for providing rapid, 
front-end polarization modulation and control over systematic errors. Despite the advantages provided by the VPM, it is 
important to identify and mitigate any time-varying effects that leak into the synchronously modulated component of the signal. 
In this paper, the effect of emission from a $300 \unit{K}$ VPM on the system performance is considered and addressed. Though instrument 
design can greatly reduce the influence of modulated VPM emission, some residual modulated signal is expected. VPM emission is treated 
in the presence of rotational misalignments and temperature variation. Simulations of time-ordered data are used
to evaluate the effect of these residual errors on the power spectrum. The analysis and modeling in this paper guides 
experimentalists on the critical aspects of observations using VPMs as front-end modulators. By implementing the characterizations 
and controls as described, front-end VPM modulation can be very powerful for mitigating $1/f$ noise in large angular scale 
polarimetric surveys. None of the systematic errors studied fundamentally limit the detection and characterization of B-modes 
on large scales for a tensor-to-scalar ratio of $r=0.01$. Indeed, $r<0.01$ is achievable with commensurately improved 
characterizations and controls.
\end{abstract}

\maketitle

\section{Introduction} \label{s:introduction}

Cosmic microwave background (CMB) measurements have been the cornerstone of cosmological advances. This is particularly true 
for measurements that test and constrain inflation. The COBE space mission discovered the anisotropy and measured the 
amplitude of scalar fluctuations \citep{Smoot1992COBE, Bennett1994COBE}. The WMAP space mission confirmed the COBE scalar amplitude 
and reported the predicted small deviation from scale invariance with the scalar spectral index, $n_s < 1$, and flatness 
to within a fraction of a percent \citep{WMAP9yrCosmo2013,WMAP9yrMaps2013}. The Planck space mission confirmed the 
WMAP results and placed tight new limits on the gaussianity of the fluctuations \citep{Planck2015Cosmo}. Looking forward, 
the next challenging step is to test and constrain inflation by characterizing the predicted stochastic background 
of gravitational waves via their imprint onto the $B$-mode polarization of the CMB. Measurement of this signal would 
establish the energy scale of inflation. It remains to be seen if characterization can be done from either the ground or a balloon, 
especially at large angular scales, or if it will require another space mission. The current best $95\%$ upper limit on $r$, the tensor-to-scalar 
ratio, is $r < 0.07$ at $k = 0.05 \unit{Mpc^{-1}}$ \citep{BICEP2KeckVI2015}.

On angular scales less than $\approx 2^{\circ}$ ($\ell > 100$), the cosmological $B$-mode angular power spectrum is dominated by 
the gravitational lensing of the curl-free $E$-mode polarization spectrum that is generated by density perturbations in the early universe 
\citep{Zaldarriaga1998,HuOkamoto2002}. At larger angular scales, $B$-modes from inflationary gravitational waves are predicted to 
source two distinct features in the angular power spectrum: an enhancement at $\ell \sim 100$ corresponding to the horizon size at 
recombination and another enhancement at low $\ell < 10$ corresponding to the size of the horizon at reionization \citep{Zaldarriaga1997,Kamionkowski1997b}. 
A compelling detection of a gravitational wave background requires the characterization of $B$-modes at angular scales $\ell < 200$. 
Such a detection would not only test the amplitude of the posited signal, but also the predicted shape of the $B$-mode angular power spectrum.

Measuring the polarization is challenging due to the minute signal level and the fact that this signature must be characterized in the presence 
of relatively large backgrounds. These backgrounds include the unpolarized radiation from the CMB, 
Galactic emission, emission from the instrument, and in the case of ground-based and balloon-borne instruments, 
emission from the atmosphere and the ground. A key aspect of CMB polarimetry is the use of temporal, spatial, or phase modulation to enable separation 
of the desired polarized signal from residual environmental influences on the instrumental response.

Modulation is the process of encoding signals of interest in a manner they can be subsequently synchronously demodulated and detected with high fidelity. Instrument 
influences outside the signal band and reference phase can be highly suppressed by the demodulation process.
The process of 
modulation is inherently a ``lock-in'' measurement, and its use in microwave radiometry dates back to \citet{Dicke1946}. 
In this signal processing technique, a modest degradation in sensitivity is traded for a dramatically increased signal stability and/or control over environmental effects.

A well-designed instrument and observing strategy commonly employs multiple levels of modulation at differing frequency to aid in the separation of the 
desired signal from spurious artifacts. These levels are distinguished by their characteristic 
time scales. Generally, the primary modulation, the one implemented on the shortest time scales, is critical because it
sets the instrument's ability to mitigate variations in the instrument and environment. Such variations tend to be more 
severe at low frequencies and are typically characterized by a $1/f^{\alpha}$ functional form, where $\alpha$ is the spectral index of 
the variations. The primary modulation frequency 
is chosen to be above the knee frequency of the $1/f^{\alpha}$ spectrum. By only measuring the signal at the modulation 
frequency and phase (by demodulating and detecting it), the instrument limits noise contributions outside of this band and 
specifically suppresses the effect of the variations in the system and environment that occur at lower frequencies than 
that of the modulation. Secondary modulation strategies are generally employed to remove systematic effects from the 
targeted signal. For example, instrument rotation can be employed to separate an astronomical polarization that is fixed in 
sky coordinates from an instrumental one that is constant in instrument coordinates
\citep{POLAR2003,HinderksQUaD2009,BICEP3yr2014,BICEP2II2014,Keck2015}.

The method of primary modulation most commonly utilized in CMB polarization experiments is an evolution of that developed in the context of 
continuum radiometry and photometry \citep{Emerson1979,Emerson1988,Magnum2007}. These approaches have been successully applied in 
mapping the CMB anisotropy -- the telescope is scanned in angle over a patch on the sky such that spatial frequencies are modulated 
and subsequently encoded at temporal frequencies in the data 
\citep{Wright1996,Tegmark1997,PrykeQUaD2009,StorySPT2013,DunnerACT2013,WMAP9yrMaps2013,BICEP22014,PB2014BB}. 
Limits on telescope capabilities and variations in the \textit{unpolarized} emission from the sky over large angular scales have typically 
limited suborbital scanning experiments to multipoles corresponding to the recombination feature and higher.

An alternative to spatial scan modulation is to directly encode the incident polarization signal. 
System level implementations of polarization modulation employ such devices such as half-wave plates (HWP)
\citep{EBEx2010SPIE,Spider2013JCAP,ABS_HWP_2013}, Faraday rotators \citep{Moyerman2013}, and phase switched receiver architectures 
\citep{WMAPRadiometers2003,PlanckLFI2010,QUIETInstrument2013}
that vary the polarization state of the 
incoming radiation for synchronous demodulation and detection of the signal.
In the ideal case, unpolarized and polarized components of the incoming radiation are treated independently by the system 
architecture and analysis pipeline. This approach enables configurations that approximate 
a null result when observing an unpolarized source. This is of particular value for mitigation of the effectively unpolarized atmospheric 
emission on large angular scales.
However, the utility of a given modulation technique ultimately depends on how well one can limit non-ideal performance during observations and calibration 
and, if necessary, model and remove such behavior via an unbiased procedure to achieve the desired measurement precision.

VPMs operate by introducing a variable phase delay between two orthogonal linear polarizations 
(see \autoref{fig:VPMdiagram} and \cite{ChussVPM2006}). 
Such modulators are implemented by placing a polarizing grid in front of and parallel to a flat mirror. The phase, and 
therefore the polarization sensitivity of the instrument, is controlled by varying the separation between the grid and the mirror. VPMs are being 
applied to the measurements of the CMB polarization at large scales. This has many advantages. First, VPMs can be made large 
enough to be usable as the first optical element for meter-sized telescopes. This minimizes systematic errors due to temperature 
to polarization leakage in the instrument.
Second, for CMB wavelengths and meter-sized apertures,
the modulation can be physically implemented at a frequency of several Hz, which moves the signal above the typical knee frequency of the sky noise for ground-based 
observations. Additionally, the VPM allows independent measurements of polarization at 
each beam position without relying on scanning to reject the excess low-frequency noise. Fast polarization modulation can enable map-making without 
first differencing time-ordered data (TOD) from co-pointed orthogonally polarized detectors to remove the unpolarized signal. 
VPMs, due to the fact that they modulate polarization with linear-circular conversion, 
reduce signal contamination due to cross-polarization ($Q \leftrightarrow U$) compared to wave plates \citep{Chuss2010SPIE}.
Since the circular polarization in the CMB is expected to be low \citep{Kosowsky1996}, residual mixing between the linear 
and circular terms in the VPM transfer function will not cause significant systematic contamination of the linearly polarized 
CMB maps.

\begin{figure*}
\begin{center}
\begin{tabular}{ccc}
\includegraphics[width=0.33\textwidth]{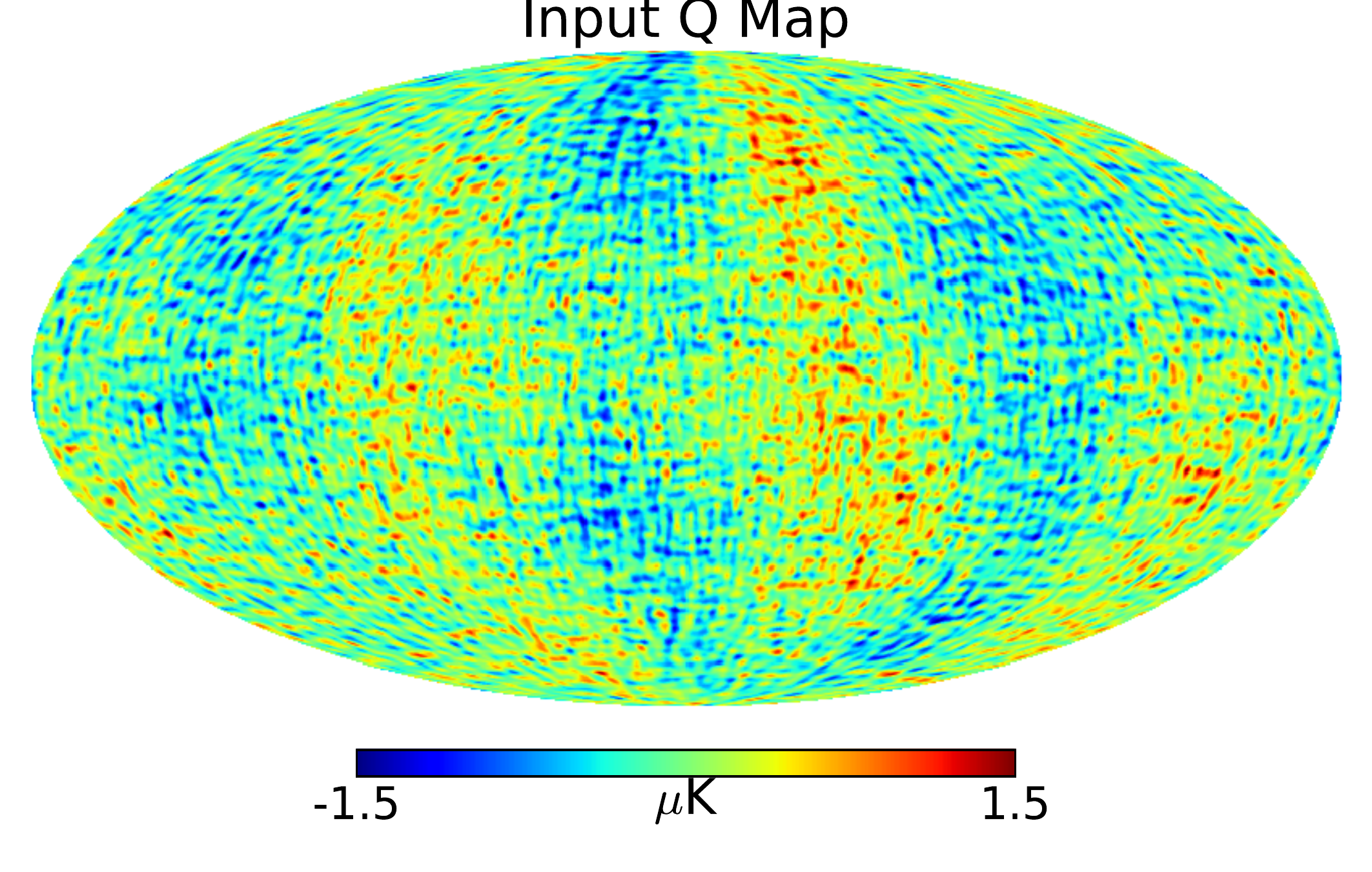} &
\includegraphics[width=0.33\textwidth]{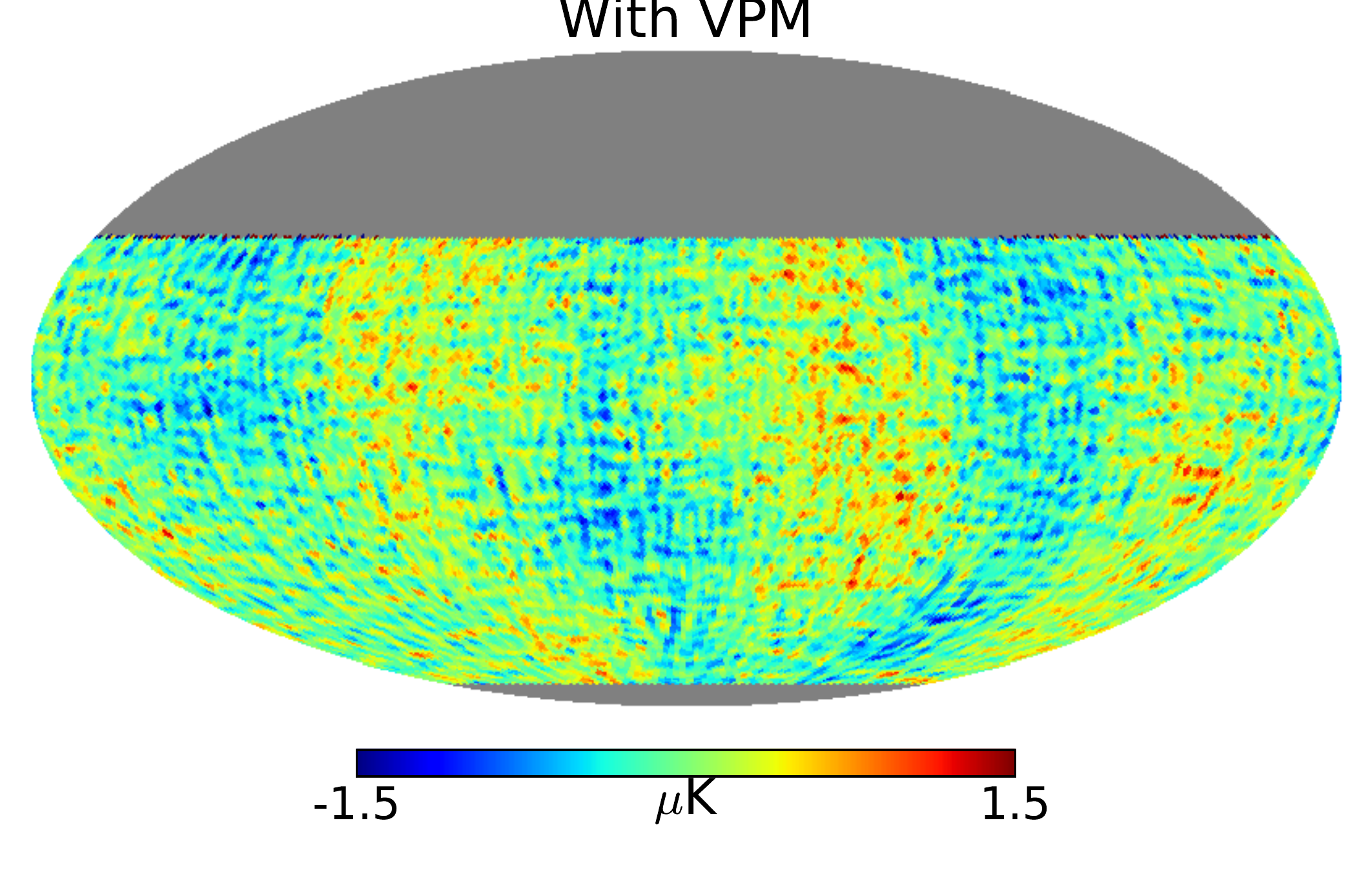} &
\includegraphics[width=0.33\textwidth]{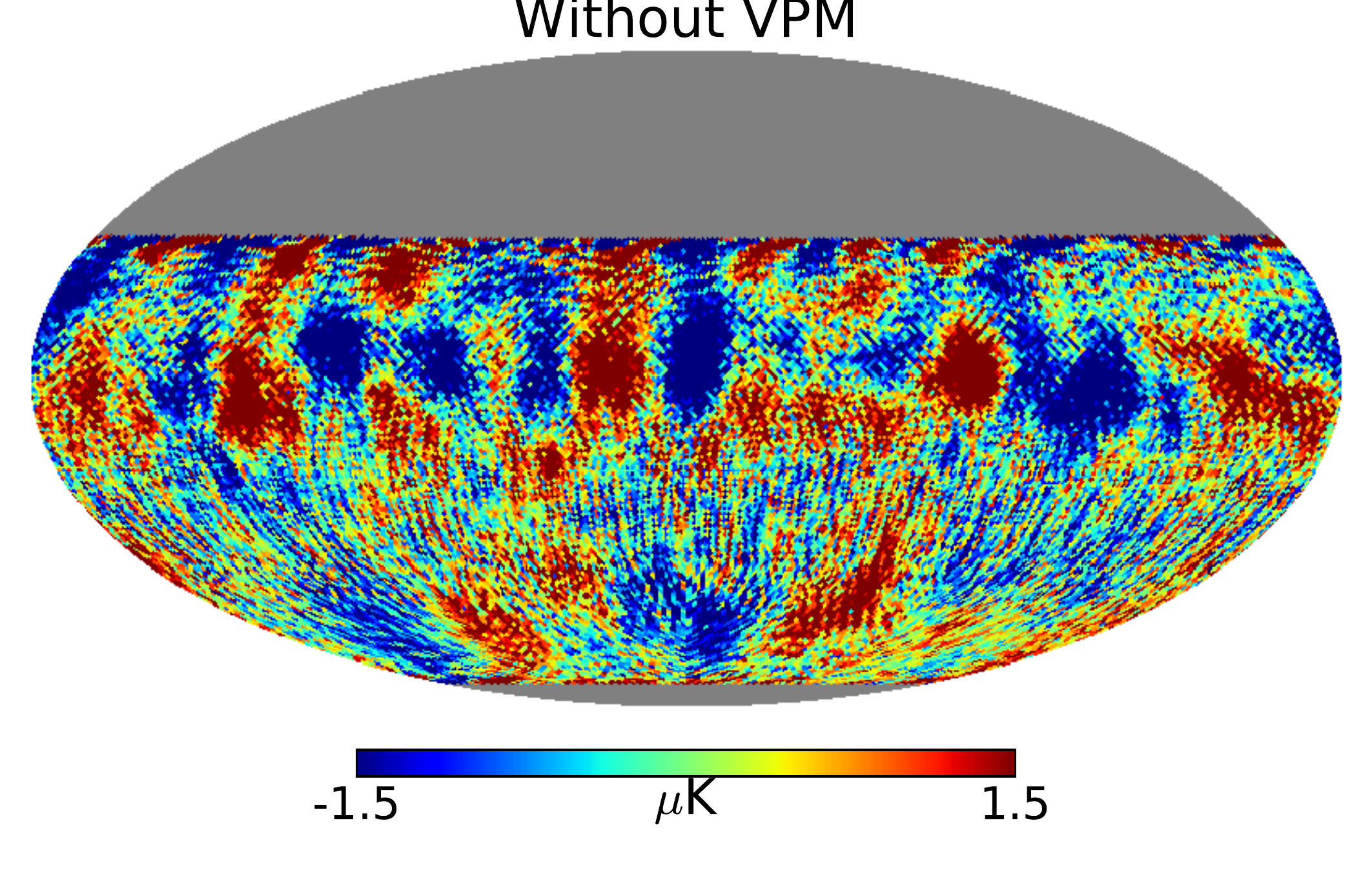}
\end{tabular}
\caption{Recovered map of Stokes $Q$ for simulations with polarized $1/f$ noise. The simulations and reconstruction follow 
the description in \autoref{s:simulations}. The amplitude of the noise has been arbitrarily chosen to illustrate the
effectiveness of polarization modulation for large scale maps. A detailed study that employs a more realistic noise level is 
treated in \autoref{s:simulations}. The leftmost panel shows the input Stokes $Q$ map.}
\label{fig:sys_modulation}
\end{center}
\end{figure*}

The advantages of a VPM for recovery of polarization at large angular scales in the presence of atmospheric variability 
are captured in the maps of \autoref{fig:sys_modulation}, which are generated following the simulation techniques discussed 
in \autoref{s:simulations}. The only difference between the simulations resulting in the two plots is the presence of a 
VPM, which modulates the sky signal before the addition of the polarized $1/f$ noise. Without a VPM, the $1/f$ noise 
contaminates the reconstructed map. With a VPM, the modulation allows the sky signal to be recovered on large angular scales. 

Ideally, a polarimeter would return a null demodulated signal or ``offset'' when viewing an unpolarized source. In reality, 
a measurement system has a finite instrumental and environmental contributions to response which in practice one strives 
to minimize their spatial-temporal variation. These ``systematic errors'' must be characterized and removed in order 
to meaningfully interpret the data. This is especially important in CMB polarization measurements for which the signal 
of interest is small, extended, and measured in the presence of large backgrounds. 
Detailed studies of the instrumental response have been previously carried out for a variety of system architectures 
\citep{Shimon2008,MacTavish2008,Miller2009,Karakci2013,ABS_HWP_2013}. 
This paper undertakes a similar analysis for a VPM-based polarimeter architecture. In this work we focus on systematic 
influences arising from the radiative properties of the modulator, and develop techniques for separating these instrumental 
artifacts from the desired astrophysical signal.

VPMs will be deployed on two upcoming CMB experiments. The first experiment is the Cosmology Large Angular Scale Surveyor (CLASS).
CLASS is currently being deployed to the Atacama desert in Chile \citep{CLASS2014SPIE}. 
CLASS will consist of one $38 \unit{GHz}$ 
telescope, two $93 \unit{GHz}$ telescopes, and one dichroic $148/217 \unit{GHz}$ telescope. Each telescope will have a VPM at 
$\sim 280 \unit{K}$ as its first optical element and 
they will observe $\sim 70\%$ of the sky from their $-23^{\circ}$ (south latitude) location in Chile. 
VPMs at $1.5 \unit{K}$ will be deployed on the Primordial Inflation Polarization Explorer (PIPER). PIPER is a 
balloon experiment \citep{PIPER2014SPIE} that will operate at $200$, $270$, $350$, and $600 \unit{GHz}$.

This work uses simulations to propagate a realistic model of the instrument's response in the time domain through polarized
power spectra. The structure of this paper is as follows. \autoref{s:vpmmodel} gives an overview of the effect of the VPM 
on the input signal. \autoref{s:systematics} introduces systematic effects generated by the VPM. 
\autoref{s:simulations} describes the simulations we perform. \autoref{s:results} shows the 
results of those simulations. Finally, we give the conclusions in \autoref{s:conclusions}.

\section{VPM Modulation} \label{s:vpmmodel}

\begin{figure}
\begin{center}
\includegraphics[width=0.7\columnwidth]{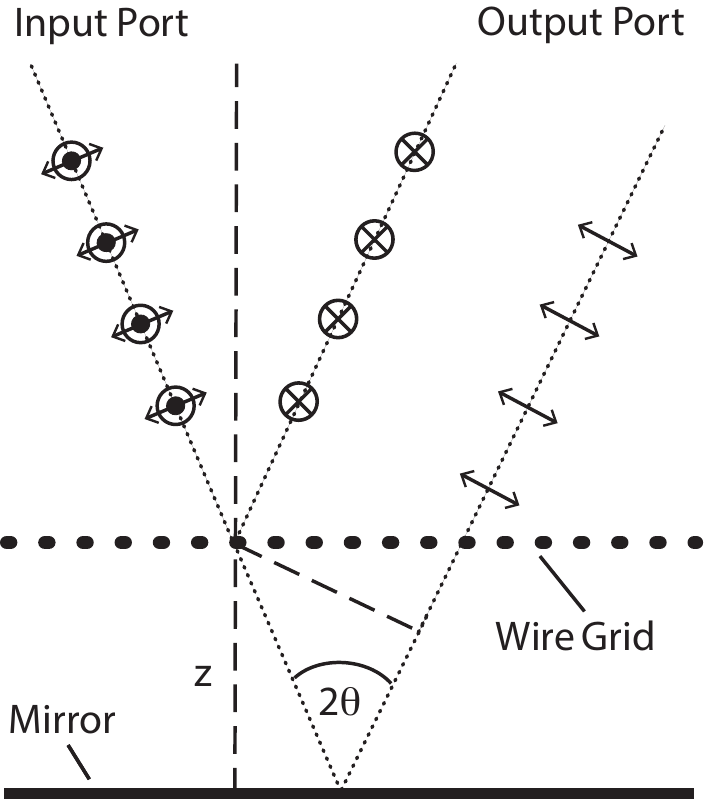}
\caption{A diagram of the VPM. Light that is polarized parallel to the grid wires is reflected. Light that is polarized 
perpendicular to the wires is transmitted through the grid and reflected off of the mirror. The two polarizations are recombined at the 
output with a relative phase delay that is a function of the grid-mirror physical separation, $z$. By modulating $z$, the polarization state 
is modulated. For CMB frequencies, $z$ will range from $\approx 0.5 - 5 \unit{mm}$.}
\label{fig:VPMdiagram}
\end{center}
\end{figure}

The VPM modulates polarization by introducing a controlled, variable phase delay between orthogonal linear polarization states. 
As shown in \autoref{fig:VPMdiagram}, a common implementation of a VPM consists of a wire grid polarizer placed in front of and 
parallel to a translating mirror. Light polarized parallel to the grid wires reflects off of 
the polarizer; light that is polarized perpendicular to the wires is transmitted through the grid and reflected off of the mirror. The two polarizations are recombined 
at the output with a relative phase delay that is proportional to the grid-mirror physical separation, $z$.

One can visualize the effect of this introduced phase delay as follows. When the phase delay is zero, the output polarization is unchanged. 
As the phase delay is increased to $\pi/2$, the VPM becomes a reflective quarter-wave plate, converting linear polarization to circular 
polarization. Increasing the phase delay further to $\pi$, one recovers the behavior of a reflective HWP for which the direction of 
linear polarization is reflected about the line parallel to the wires.

\subsection{An Ideal VPM} \label{s:vpmidealmodel}

The key characteristic of an ideal VPM is that each polarization is treated by the VPM independently with $100\%$ 
transmission for the light polarized perpendicular to the wires and $100\%$ reflection for the light polarized parallel to the wires. Thus, 
the ideal system is lossless with perfect polarization isolation.
Defining a coordinate system such that the detectors are sensitive to Stokes $Q$, we assume an architecture in which the grid wires 
are at $\pm 45^{\circ}$ with respect to the orientation of the detectors' polarization 
sensitivity. The polarization transformation of the VPM for monochromatic light can be written as 
\begin{equation}
    \mathbf{s}' = \mathbf{M} \mathbf{s}
\end{equation} where $\mathbf{M}$ 
is the Mueller matrix and $\mathbf{s}$ and $\mathbf{s^\prime}$ are the vectors of Stokes parameters before and after the transformation, 
respectively. Writing out the vectors and matrices gives us:
\begin{equation} \label{eq:vpmmueller}
\left( \begin{array}{c} I' \\ Q' \\ U' \\ V' \end{array} \right) = \left( \begin{array}{cccc} 1 & 0 & 0 & 0 \\ 0 & \cos \phi & 0 & - \sin \phi \\
0 & 0 & 1 & 0 \\ 0 & \sin \phi & 0 & \cos \phi \end{array} \right) \left( \begin{array}{c} I \\ Q \\ U \\ V \end{array} \right)
\end{equation}
Here, $\phi = \phi(z)$ is the phase delay and $z$ is the grid-mirror separation. The primed and unprimed Stokes parameters represent 
the polarization states on the input and the output side of the VPM, respectively. 
For the case in which the wavelength is much larger than the wire diameter, the phase delay can be approximated as proportional to the 
path length difference \citep{Chuss2012}. 
\begin{equation} \label{eq:phaserelation}
\phi(z) \simeq \frac{4 \pi z}{\lambda} \cos \theta
\end{equation}
where $\lambda$ is the wavelength of the radiation and $\theta$ is its incident angle on the VPM. 

As can be seen from \autoref{eq:vpmmueller}, the VPM modulates one of the linear Stokes parameters (in this case, $Q$) in addition to circular 
polarization (Stokes $V$). This $Q \leftrightarrow V$ modulation is due to the $45^\circ$ orientation of the grid wires with respect to the detectors.
If the grid wires are instead at $0^{\circ}$ or $90^{\circ}$ in this 
coordinate system, the VPM would produce $U \leftrightarrow V$ modulation, and has no influence of the $Q$ signal measured by the signal 
difference of an
orthogonal pair of linearly polarized sensors.

By using sky and/or instrument rotation, the modulation of Stokes $Q$ in the 
instrument frame can be used to measure both Stokes $Q$ and $U$ on the sky. The tracking of instrument rotation and pointing position of the instrument will
also be utilized as secondary modulation schemes to track and remove linear polarization from ground spill and instrument emission.
Because the VPM transforms circular polarization to linear polarization, VPM modulation can be used to construct a map of Stokes $V$ in 
addition to $I$, $Q$, and $U$. For the CMB, the theoretical expectation is that the circular polarization will be negligible 
compared to the linear polarization \citep{Kamionkowski1997b,Cooray2003,Alexander2009}. 
There will be additional circular-polarized contributions from the atmosphere \citep{Keating1998,HananyRosenkranz2003,Spinelli2011,Errard2015}
that are not spatially uniform. As such, the Stokes $V$ channel will likely be utilized to track these spurious systematic effects, and the sensitivity of the architecture
to astronomical circular polarization will depend on the magnitude of such effects.

\subsection{Transmission-Line Model} \label{s:vpmtlmodel}

To capture details of the VPM operation beyond the ideal model, we implement a transmission-line model for the VPM 
based on \cite{Chuss2012}. The two main manifestations of the improved fidelity of the transmission-line model 
over the ideal model are: (1) a small modification of the distance/phase relation in \autoref{eq:phaserelation}; and (2) precise 
treatments of the polarization isolation and finite absorptance of the VPM's components. This enables the estimation 
of the differential emissivity between the polarization parallel and perpendicular to the wires, which is a key source of 
systematic effects studied here.

The responses of the wire grid polarizer to parallel and perpendicularly polarized incident radiation are modeled via 
inductive and capacitive circuits, respectively. Loss in the parallel component is included via a series resistance that is calculable 
from the wire geometry and material. Loss in the perpendicular component can be treated in an analogous way, but because it is 
sub-dominant it is not included in this model. 

A combination of the finite conductivity of the wires and the imperfect isolation between the reflected and transmitted polarizations leads to a higher 
emissivity for the parallel than the perpendicular polarization. In addition to emission due to its finite physical temperature, the VPM 
also absorbs a small fraction of the signal from the sky. 
The overall instrumental signal is proportional to the differential emissivity of the wires and the product of the temperature difference between the 
physical temperature of the VPM and the brightness temperature of the background, which in this case consists of contributions from the atmosphere, the CMB, 
and foreground sources. It can be written as
\begin{equation} \label{eq:sVPM}
    s_{\rm VPM} = \left[T_{\rm VPM} - \left(\epsilon_{\rm atm} T_{\rm atm}+T_{\rm cmb} +\sum_i \epsilon_i T_i\right)\right] \epsilon_{\rm VPM}
\end{equation}
Here, $T_{\rm VPM}$, 
$T_{\rm atm}$, and $T_{\rm cmb}$ are the physical temperatures 
of the VPM, atmosphere and CMB, respectively. $T_i$ and $\epsilon_i$ are the temperature and emissivity of any astrophysical foregrounds in the instrument's beam. 
$\epsilon_{\rm VPM}$ is the differential emissivity of the VPM. The result, $s_{\rm VPM}$, is the systematic signal emitted from the VPM.
In this work, we consider the CMB as the only celestial source, so the terms in the summation are set to be zero. The magnitude of this systematic signal is a 
function of grid-mirror position.
Specifically, the differential emissivity component, $\epsilon_{\rm VPM} (z)$ (and consequently the systematic signal) is a function of grid-mirror position. 
For a warm VPM ($T_{\rm VPM} > (\epsilon_{\rm atm} T_{\rm atm}+T_{\rm cmb} )$), this will lead to a modulated grid emission. For a cold VPM 
($T_{\rm VPM} < (\epsilon_{\rm atm} T_{\rm atm}+T_{\rm cmb} )$), this will lead to a modulated grid absorption. 

\begin{figure}
\begin{center}
\includegraphics[width=1.0\columnwidth]{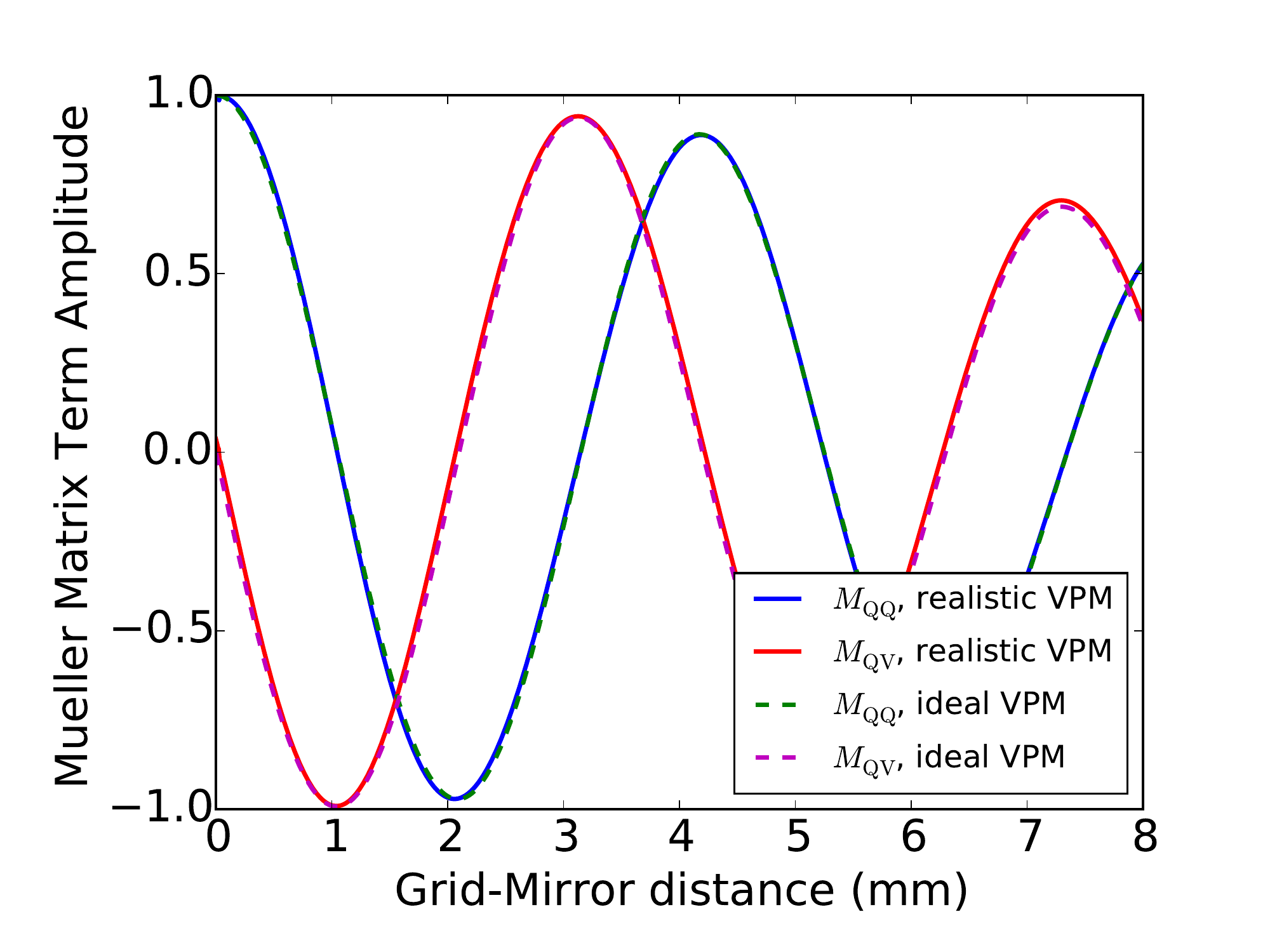}
\caption{Mueller matrix terms for a VPM described by a transmission-line (solid lines) and an ideal (dashed lines) model bandpass averaged 
from $33$ to $43 \unit{GHz}$ with the wire grid at $45^{\circ}$ with respect to the polarization sensitive direction of the detectors. 
The blue and green lines are the $M_{\rm QQ}$ terms, while the red and magenta lines are the $M_{\rm QV}$ terms. 
The decrease in amplitude at larger grid-mirror separations results from the decoherence in the modulation across the frequency band of the detectors.}
\label{fig:VPMmodulation}
\end{center}
\end{figure}

\begin{figure}
\begin{center}
\includegraphics[width=1.0\columnwidth]{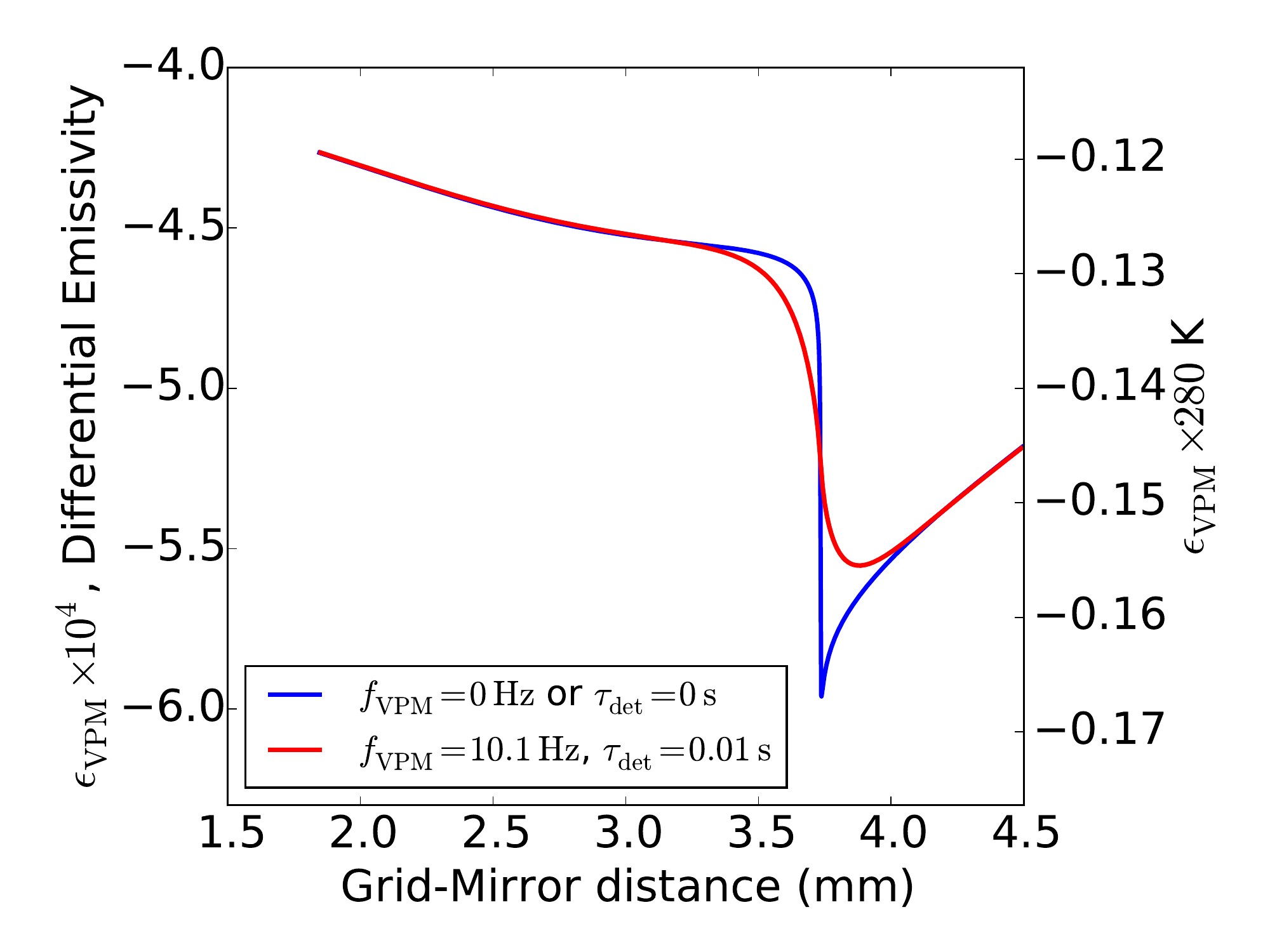}
\caption{Differential emissivity of the VPM as a function of grid-mirror separation band averaged over $33$ to $43 \unit{GHz}$. 
This signal leaks into the local $Q$ and $U$ depending on the orientation of the grid. The blue line assumes either the detector time 
constant, $\tau_{\rm det}$, or, equivalently, the modulation frequency of the VPM, $f_{\rm VPM}$, is zero. The red line is 
the effective (smoothed) differential emissivity for the case in which the time constant of the detectors is $0.01 \unit{s}$ and
the modulation frequency of the VPM is $10.1 \unit{Hz}$. This is representative of a practical system implementation.}
\label{fig:diffemissivity}
\end{center}
\end{figure}

Selected band-averaged Mueller matrix elements for the VPM transmission line model that has been implemented for this work are 
shown in \autoref{fig:VPMmodulation} superposed on the corresponding elements for the ideal model. We have implemented the 
transmission line model described by \cite{Chuss2012} using a grid having $50 \unit{\mu m}$ diameter wires on a $110 \unit{\mu m}$ 
center-to-center wire spacing. The surface of the wire is copper with a conductivity of $6\times10^{7} \unit{\mathrm{S}\cdot \mathrm{m}^{-1}}$.
Differential emissivity is shown in \autoref{fig:diffemissivity}. 
A resonant feature in the differential emissivity occurs due to the presence of a cavity resonance defined by the grid-mirror separation. This resonance 
is associated with the mode having its polarization parallel to the wires being trapped 
by the inductive grid.

The broadband spectral response of the detectors through the VPM is assumed to be a top hat function. The 
frequencies of operation are taken to be between $33$ and $43 \unit{GHz}$ ($9$ to $7 \unit{mm}$), corresponding to the low frequency 
channel of CLASS \citep{CLASS2012SPIE,CLASS2014SPIE}. In the simulation, the frequencies are averaged with a spacing of $0.002 \unit{GHz}$. The 
feature observed in \autoref{fig:diffemissivity} at $\sim 3.7 \unit{mm}$ is associated with the resonant VPM response in the detector band.  As verification of 
this physical interpretation, we find this resonance disappears when the inductive cavity is removed and the polarization parallel to the wires 
is no longer trapped. We verified that the frequency resolution stated above was sufficient to Nyquist sample the VPM response at all relevant 
time scales. For a real system, the bandpass and transfer function of the VPM will need to be characterized, but a top-hat frequency response 
is sufficient to capture the critical aspects of the system response.

\section{Systematics} \label{s:systematics}

VPM-based effects, such as VPM emission, are naturally separated from the 
signal in each detector in the following way. As the grid-mirror separation is varied, the VPM will differentially affect the linear polarizations 
parallel and perpendicular to its wires. Because the wires are at a $45^{\circ}$ angle to the polarization sensitive direction of the detectors, such linear 
polarized signals will only be present in the local Stokes $U$ parameter, while each detector is only sensitive to the local Stokes $Q$ parameter. 
The CMB signal, on the other hand, will be present in the local Stokes $I$, $Q$ and $U$ parameters. In this section, we describe how 
deviations from this symmetry can couple artifacts into the signal.

\subsection{Grid Misalignment Error}

We first examine the effect of a grid misalignment error. 
Misalignment arises through an error in the assumed angle between 
the VPM grid wires and the polarization axis of the detectors. 

\begin{figure}
\begin{center}
\includegraphics[width=1.0\columnwidth]{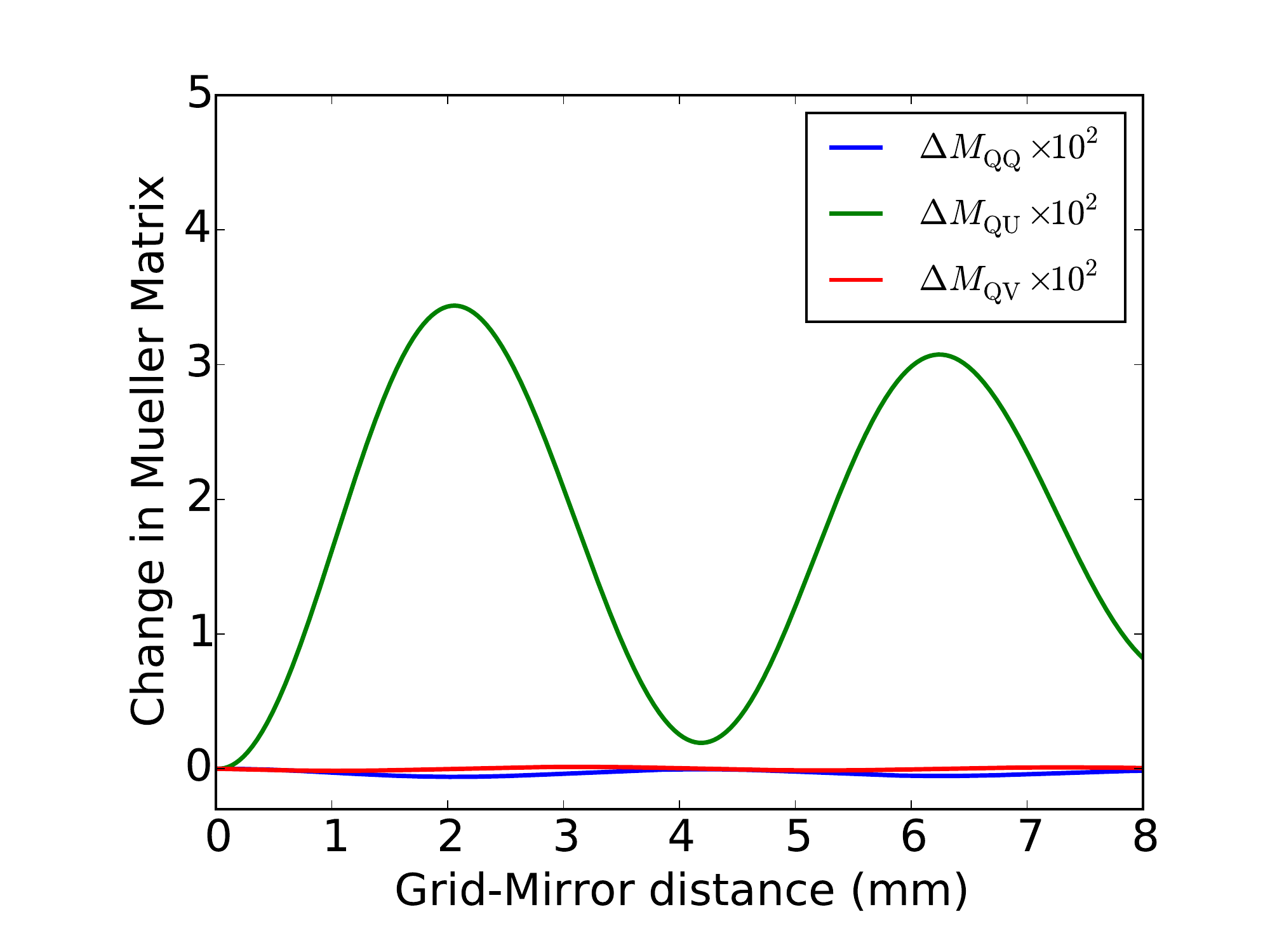}
\caption{Difference in the Mueller matrix terms between the cases for the grid-detector angle of $45^{\circ}$ and $44.5^{\circ}$. Results are for 
$M_{\rm QQ}$ (blue), $M_{\rm QU}$ (green), $M_{\rm QV}$ (red).}
\label{fig:gridangleerror}
\end{center}
\end{figure}

The Mueller matrix of the system is dependent on the grid-detector angle. For example, with a perfect $45^{\circ}$ alignment, $M_{QQ}$ and $M_{QV}$ 
are non-zero while $M_{QU}$ is zero. As 
the angle alignment changes, $M_{QU}$ becomes non-zero, but is still small compared to $M_{QQ}$ and $M_{QV}$ for small deviations from $45^{\circ}$. This is 
illustrated in \autoref{fig:gridangleerror} which shows the difference in Mueller matrix terms between grid-detector angles of $45^{\circ}$ and $44.5^{\circ}$. 
We test the sensitivty of the maps and associated power spectra to grid misalignment error.

\subsection{VPM Emission and Absorption}

The main set of systematics that we address in this work are those arising from the coupling of the VPM emission into the signal. As 
described in \autoref{s:vpmtlmodel}, the transmission-line model for the VPM includes the effect of the differential emissivity 
of the wires, which can convert an unpolarized signal into a polarized signal. Since the light is emitted, absorbed, and polarized 
preferentially along the direction of the grid wires, these systematics are balanced by placing the grid wires 
at $45^{\circ}$ with respect to the polarization direction of the detectors. In this case, in the coordinate system defined by the 
polarization directions of the detectors (the detector frame), the VPM emission 
and absorption would only affect the non-measured $U$ Stokes parameter (i.e. the Mueller matrix term describing $I$ to $Q$ leakage, $M_{QI}$, would be zero). 
Under the condition of perfect $45^{\circ}$ alignment between the grid wires and the VPM, the variation in the $I$ signal in each detector pair 
is common mode. In contrast, the $Q$ signal is \textit{differential} in the two detectors.

There will always be uncertainty in the orientations of the grid wires, and thus they will not be at exactly $45^{\circ}$ with respect to the detectors. 
If this relative angle differs from $45^{\circ}$ by an angle $\delta$, $M_{QI}$ would be non-zero and 
there will be a VPM emission signal of 
\begin{equation} \label{eq:gridemission}
    s_{\rm ve} = s_{\rm VPM} \sin \delta 
\end{equation}
leaking to $Q$ in the detector frame, where $s_{\rm VPM}$ is the total amplitude of the systematic signal emitted by the VPM 
and given by \autoref{eq:sVPM}. 
The physical temperatures can be inferred from the data, while $\epsilon_{\rm VPM}(z)$, and $\delta$ 
are slowly varying and need to be measured.
With a differential emissivity of 
$1.0 \times 10^{-4}$, a temperature difference of $250 \unit{K}$ ($T_{\rm VPM} = 280 \unit{K}$ and $T_{\rm sky} = 30 \unit{K}$), 
and $\delta = 0.05^{\circ}$, the output $Q$ in the detector 
frame due to VPM emission is $2.2 \times 10^{-5} \unit{K}$. This is close to two orders of magnitude larger than the RMS of the $Q$ and 
$U$ CMB polarization signal of $\approx 3.6 \times 10^{-7} \unit{K}$, so care must be taken in separating the synchronous signature 
from the true sky signal. We discuss this in more detail in \autoref{s:simulations}.

As seen in \autoref{eq:gridemission}, the VPM emissivity amplitude is proportional to the difference between the background brightness temperature and the physical temperature  of the VPM. 
Cooling the VPM to match the brightness temperature of the atmosphere will greatly reduce the amplitude of the VPM emissivity; however, there is a trade-off in 
doing so that is similar to that of other modulators (e.g. HWPs). For a ground-based implementation, a cold modulator is required to be 
placed behind a pressure window, and thus, polarization effects of the window must be considered. 
For space-borne instruments (and suborbital platforms such as PIPER), the pressure window requirement can be relaxed and a cold modulator is more pratical.

In a real VPM-based instrument, the temperature difference in \autoref{eq:gridemission} will vary over time scales that are long 
compared to that of a single stroke of the VPM. If the grid is not at $45^{\circ}$ with respect to the polarization sensitive direction of 
the detectors, this temperature variation can introduce noise in the polarized power spectra. 

\section{Simulations} \label{s:simulations}

To investigate the mitigation of the two systematics described above, we constructed a simulation of the instrument response during observation. 
We simulate a CMB sky with the parameters from a model consistent with a Planck $\Lambda$CDM CMB model, assuming $r=0$ \citep{Planck2015Cosmo}.  We generate a simulated 
TOD by smoothing the simulated sky with a $1.5^{\circ}$ Gaussian beam and implementing an observing strategy.
Next, we generate maps from these TOD. 
Finally, we construct the power spectra of the maps and subtract the power spectra 
of simulations without instrumental effects to obtain an estimate of the systematic error power spectra. 

The only sources of noise that enter the simulation are those from reconstructing 
the template and from the temporal variations in VPM emission signature (see \autoref{s:gridemission} and \autoref{s:gridemissionTvary}).
Similarly, we ignore the sun, foregrounds, and any loss of 
observation time due to telescope maintenance or calibration to simplify the calculations.

\begin{figure}
\begin{center}
\includegraphics[width=1.0\columnwidth]{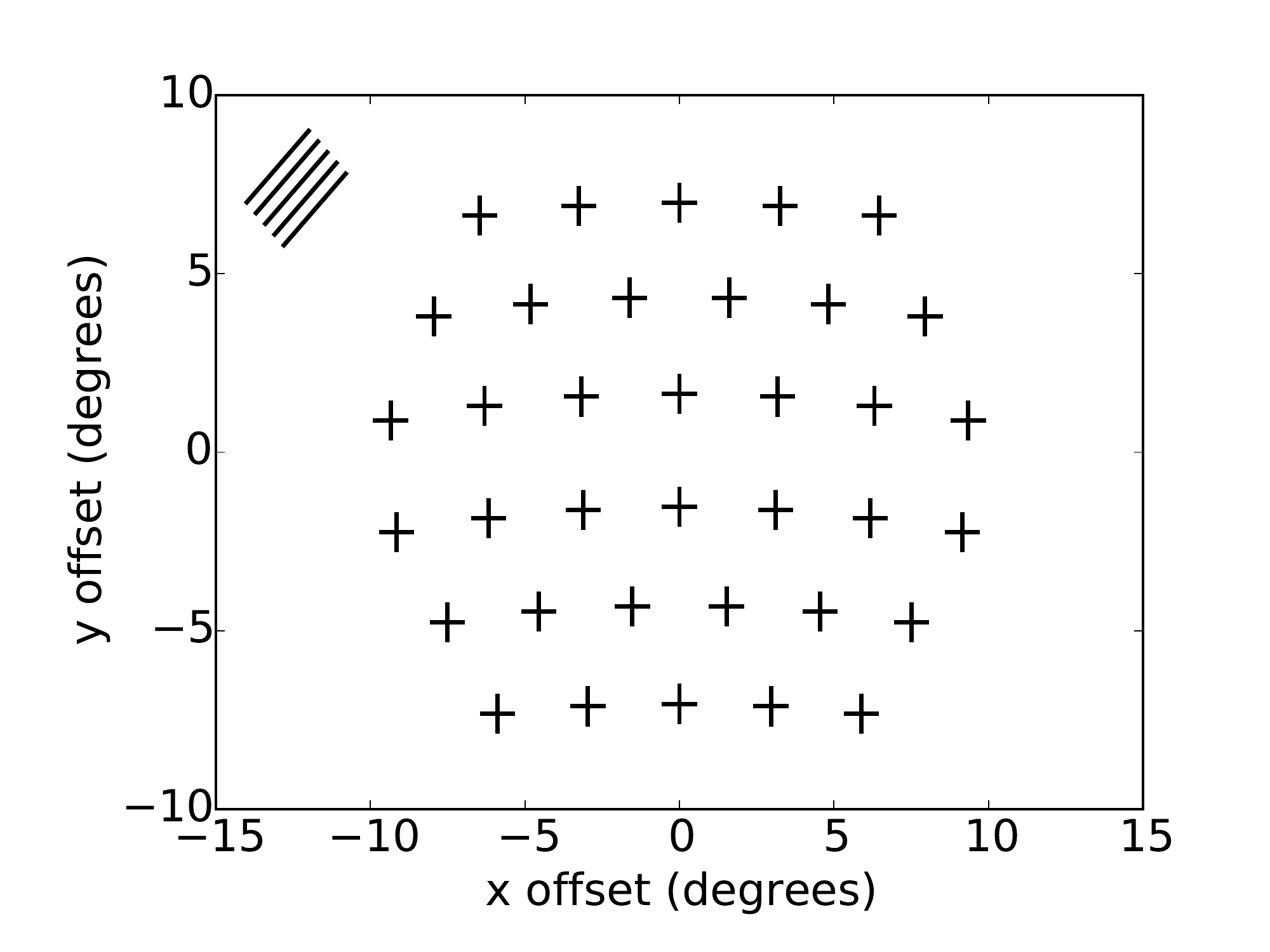}
\caption{The focal plane used in the simulations. The lines show the polarization directions of each detector used in the simulations. There are $36$ pixels 
containing $72$ polarization sensitive detectors. The lines in the upper left corner show the ideal direction that the grid makes relative to 
the polarization sensitive direction of the detectors.}
\label{fig:CLASS40GHzFocalPlane}
\end{center}
\end{figure}

To determine the effect of each systematic error term on the resulting power spectra, we run several simulations with these effects present.
As a representative test case, the simulations are modeled on the CLASS experiment and observing strategy. 
The focal plane simulation consists of $36$ pixels. 
Each pixel has two detectors with orthogonal polarization orientations, which we designate $H$ and $V$ based on their 
polarization sensitive direction.
A map of the detector pointing offsets from the boresight center of the telescope and the ideal angle of the grid is shown in 
\autoref{fig:CLASS40GHzFocalPlane}. (This is similar to the $38 \unit{GHz}$ focal plane for CLASS, though the detectors and VPM polarizer in 
CLASS will be rotated by $45^{\circ}$ with respect to the directions shown here.)
The response of each detector is modeled as a single-pole filter with an integration time of $10 \unit{ms}$.
We scan $360^{\circ}$ in azimuth at $40^{\circ}$ elevation 
for $24$ hours. This generates an observed circle whose center moves in right ascension over the course of the day. The velocity of the scan is varied 
between $0.5$ and $2 \unit{deg \hspace{1pt} s^{-1}}$ the lower and upper velocity limits for CLASS
depending on the declination to make the resulting integration time per equal-area pixel as uniform as possible over the CLASS survey area. 
Each simulation consists of $7$ days of observations. 
The boresight angle of the telescope is rotated by $15^{\circ}$ every day, through a total of $90^{\circ}$ of rotation. This will reduce 
systematics due to polarized emission that is fixed in the instrument frame. To be conservative, 
we chose these values so that, due to symmetry, the systematic error is maximized in the reconstructed $U$ maps and canceled out in the reconstructed $Q$ maps. 
This maximizes the effect of the systematic in the $BB$ power spectrum.
Our coordinate system is chosen such that at a boresight angle of $0^{\circ}$ and a parallactic angle of $0^{\circ}$, $Q$ in the detector 
frame is the same as $Q$ in the sky frame. 

For this paper, we first difference the $H$ and $V$ TOD before making maps. 
After differencing the TOD, we then solve the standard least-squares map-making equation \citep{Tegmark1997}:

\begin{equation} \label{eq:map-making}
	(\mathbf{A}^T \mathbf{N}^{-1} \mathbf{A}) \mathbf{x} = \mathbf{A}^T \mathbf{N}^{-1} \mathbf{d}
\end{equation}
where $\mathbf{N}$ is the noise covariance matrix, $\mathbf{d}$ are the measured TOD, $\mathbf{A}$ is the pointing and 
modulation encoding matrix, and $\mathbf{x}$ is a vector 
containing the Stokes parameters in each map pixel. 

The rapid polarization modulation significantly reduces the cross-coupling between pixels on the sky compared to scanning techniques. Future 
work will quantitatively address residual correlations; however, to target the specific systematic effects of interest here, we consider the noise 
to be constant and uncorrelated, allowing a useful simplication of \autoref{eq:map-making}.
If $\mathbf{N}$ is a constant multiplied by the identity matrix then $(\mathbf{A}^T \mathbf{N}^{-1} 
\mathbf{A})$ is a block diagonal matrix where there are no correlations between map pixels.
The Stokes parameters in each map pixel can therefore be solved for independently of all other pixels. 

\begin{figure}
\begin{center}
\includegraphics[width=1.0\columnwidth]{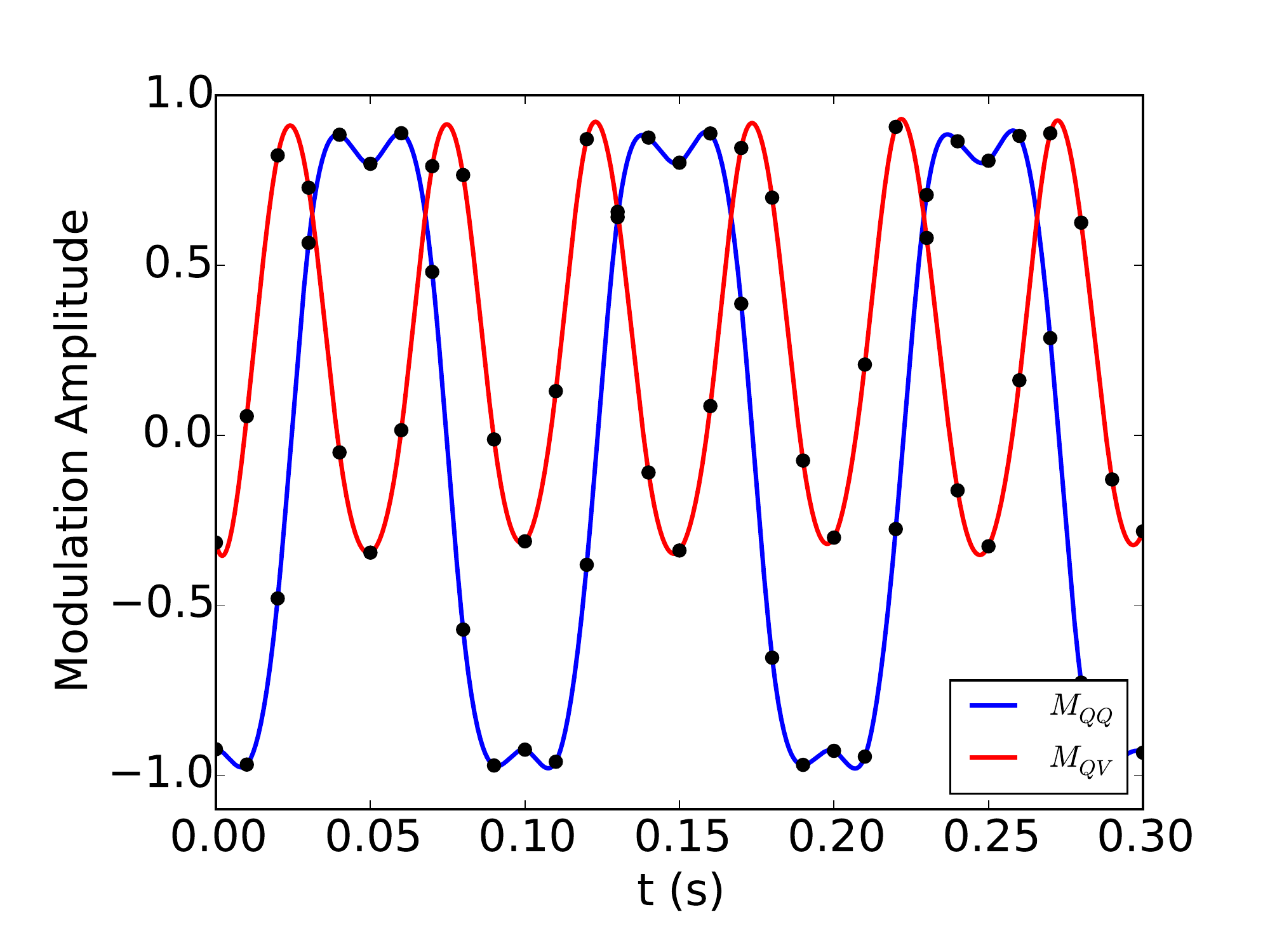}
\caption{The amplitude of the VPM modulation of the polarization signal as a function of time. The black dots are the positions sampled in the 
simulation due to the $100 \unit{Hz}$ sampling rate over the first $0.3 \unit{s}$. The VPM is modulated at a frequency of $10.1 \unit{Hz}$. The 
VPM stroke profile is chosen to maximize sensitivity to Stokes $Q$ at the expense of sensitivity to Stokes $V$.}
\label{fig:VPMmod_time}
\end{center}
\end{figure}

The pointing matrix, $\mathbf{A}$, encodes the pointing along with how the associated signal transforms as a function of the parallactic angle and VPM position. For 
an experiment with a VPM, the simplest model of the TOD, when the grid-detector angle is at $45^{\circ}$, is 
\begin{eqnarray} \label{eq:vpm_model_0}
    \mathbf{d}(z,\alpha) = Q M_{\rm QQ} \cos 2 \alpha + U M_{\rm QQ} \sin 2 \alpha + V M_{\rm QV}
\end{eqnarray}
where 
the $QQ$, and $QV$ elements of the VPM Mueller matrix ($\mathbf{M}$) are dependent on the grid-mirror separation, $z$. The parameter $\alpha$ is the sum 
of the parallactic angle and boresight angle of the telescope and is the angle between the detector coordinate system and the 
sky coordinate system. A more accurate model for the signal that is valid for general grid-detector angles is
\begin{eqnarray} \label{eq:vpm_model_1}
    \mathbf{d} &=& Q M_{\rm QQ} \cos 2 \alpha - Q M_{\rm QU} \sin 2 \alpha + U M_{\rm QQ} \sin 2 \alpha \nonumber \\ 
    &+& U M_{\rm QU} \cos 2 \alpha + V M_{\rm QV}
\end{eqnarray}
and incorporates the $M_{\rm QU}$ VPM Mueller matrix term that is non-zero when the grid-detector angle is not at $45^{\circ}$.
Plots of $M_{\rm QQ}$ and $M_{\rm QV}$ for a grid-detector angle of $45^{\circ}$ as a function of grid-mirror separation 
are shown in \autoref{fig:VPMmodulation}. The grid-mirror separation as a function of time is given by 
$z(t) = z_0 + \frac{\Delta z}{2} \left[1 - \cos (2\pi f t) \right]$, where $z_0 = 1850 \unit{\mu m}$ is the minimum grid-mirror 
separation, $\Delta z = 2640 \unit{\mu m}$ is the mirror throw, and $f = 10.1 \unit{Hz}$ is the frequency of the mirror modulation rate. 
The grid-mirror separation range is chosen to maximize signal to noise, and the modulation frequency is chosen to be slightly different 
from $10 \unit{Hz}$ to provide more than $10$ unique positions given the $100 \unit{Hz}$ sampling rate of the data. With this 
throw, we include the resonant feature in the smoothed (red) curve for differential emissivity visible in 
\autoref{fig:diffemissivity}. The corresponding transfer functions as a function of time, $M_{\rm QQ} (t)$ and $M_{\rm QV} (t)$, are 
shown in \autoref{fig:VPMmod_time}. 
The functions shown here are calculated using the transmission-line model of the VPM. In 
practice, the fidelity of a real experiment can be improved through measurement of the VPM transfer function. 
The pointing matrix, $\mathbf{A}$, encapsulates the coefficients of the three polarization signals: $Q$, $U$, 
and $V$. This technique was used in the making of Figure~\ref{fig:sys_modulation}.

After making maps, we construct the auto-correlation power spectra using standard 
{\textsc HEALPix}\footnote{\url{http://healpix.sf.net}} routines.
While we measure nearly the full sky, there is some leakage from $E$ to $B$ due to the sky cut if the standard pseudo-$C_{\ell}$ estimator 
is used. We implement the pure $C_{\ell}$ estimator \citep{KSmith2006,Grain2009} to minimize this leakage. 
The residual leakage is several orders of magnitude below $r=0.01$ in the simulations so any significant level of $B$ due to systematics 
is not associated with this reconstruction. The circular polarization $V$ is treated the same way as the temperature anisotropy when constructing 
the power spectra, because while a rotation of the coordinate system will mix $Q$ and $U$, it does not affect the measured value for either 
the $I$ or $V$ Stokes parameters. Therefore the $TT$ and $VV$ power spectra will have similar mode mixing properties when examining maps having 
partial sky coverage. 

The $C_{\ell}$ power spectra shown throughout this paper are the difference between the reconstructed power spectra with and without 
systematics averaged over $100$ realizations. This comparison is done in power spectrum space to provide sign dependence 
for the $E$-mode systematics. By differencing power spectra, we can distinguish between systematics that increase the overall power 
and systematics that might shift power from $E$-modes to $B$-modes and therefore decrease the overall amplitude of the $EE$ power spectrum.

While we show results from simulations with no input $B$-modes ($r=0$), all reconstruction techniques are also applied to simulations with a non-zero 
input $BB$ power spectrum and no systematics. We have found that the reconstruction techniques studied in this paper do not bias the underlying 
measurements, as we find that the resulting $BB$ power spectrum is unbiased compared to the input $BB$ power spectrum. The only bias is due to 
the partial sky cut and that is removed by the implementation of the pure $C_\ell$ estimator and comparisons to simulations with no systematics.
The techniques do not remove 
any measurable fraction of the input $B$-modes or convert input $E$-modes to $B$-modes.

\section{Results} \label{s:results}

We add in the systematics in three cumulative steps. First, we include the effect of grid misalignment without 
the effect of VPM emission. 
After understanding these results, we add in a non-zero VPM emission. Finally, in the last simulation we allow the temperature difference 
between the VPM and the atmosphere to vary over time.

\subsection{Grid Misalignment Error} \label{s:gaeresults}

\begin{figure*}
\begin{center}
\begin{tabular}{ccc}
\includegraphics[width=0.33\textwidth]{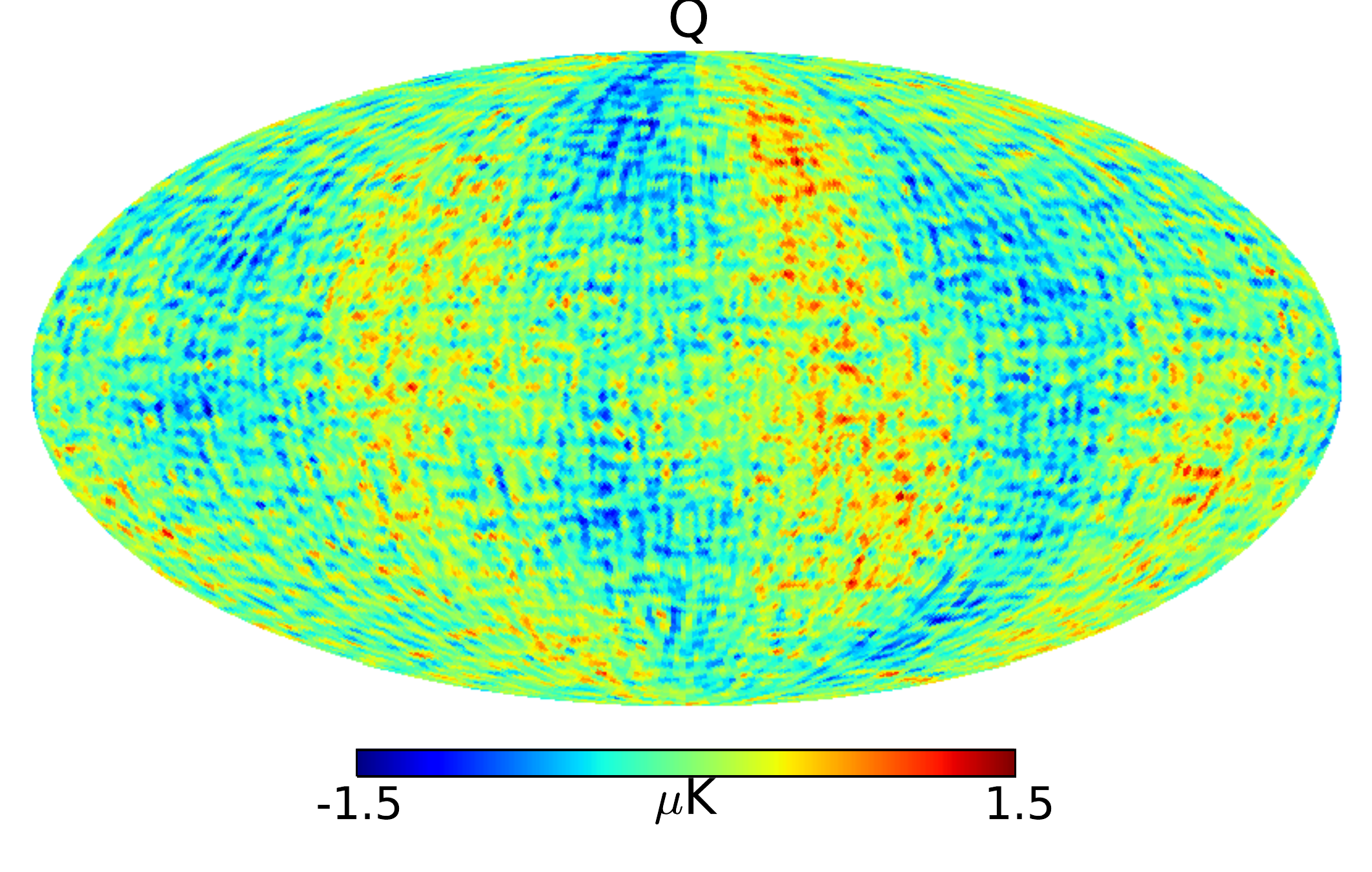} &
\includegraphics[width=0.33\textwidth]{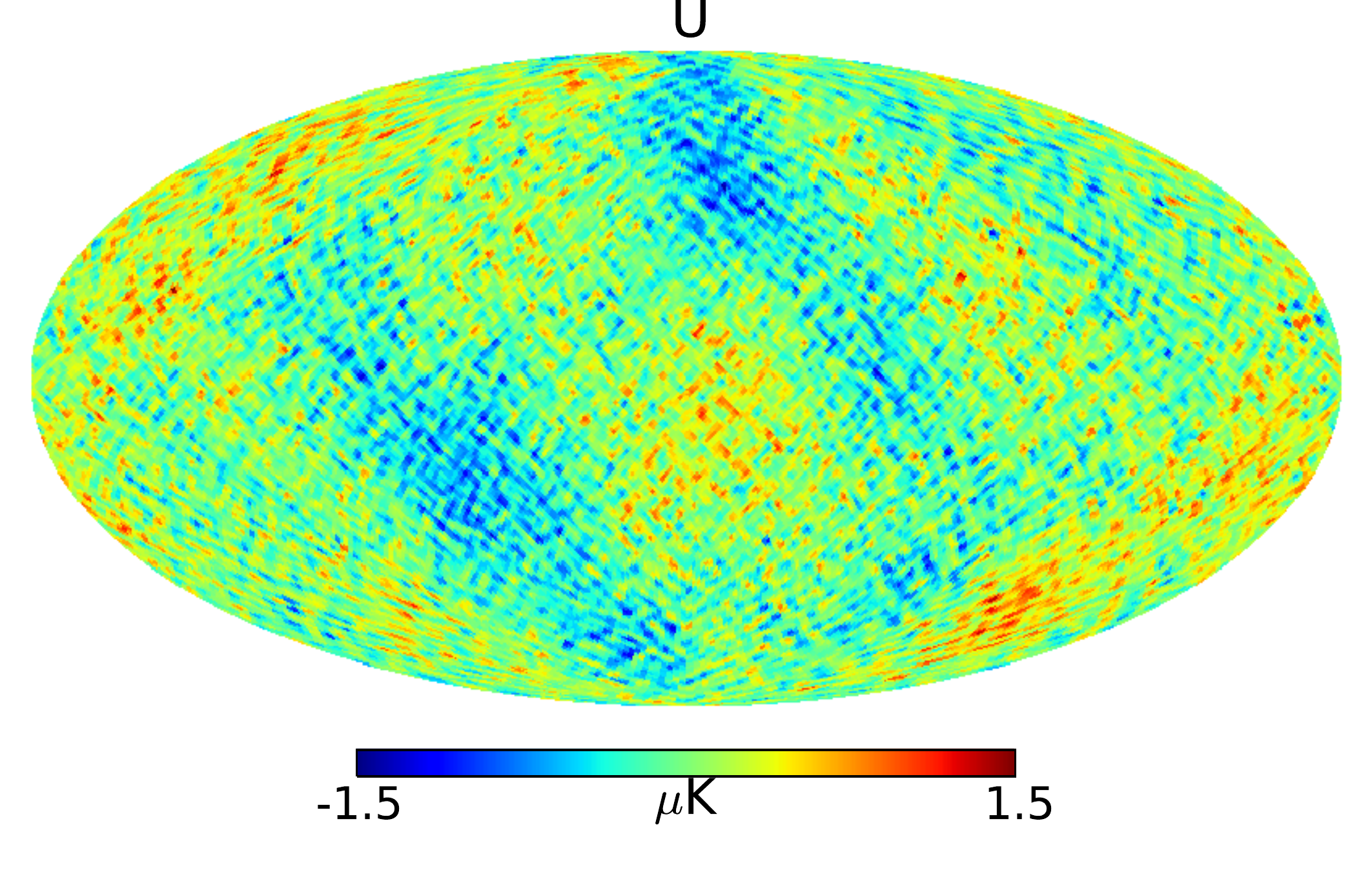} &
\includegraphics[width=0.33\textwidth]{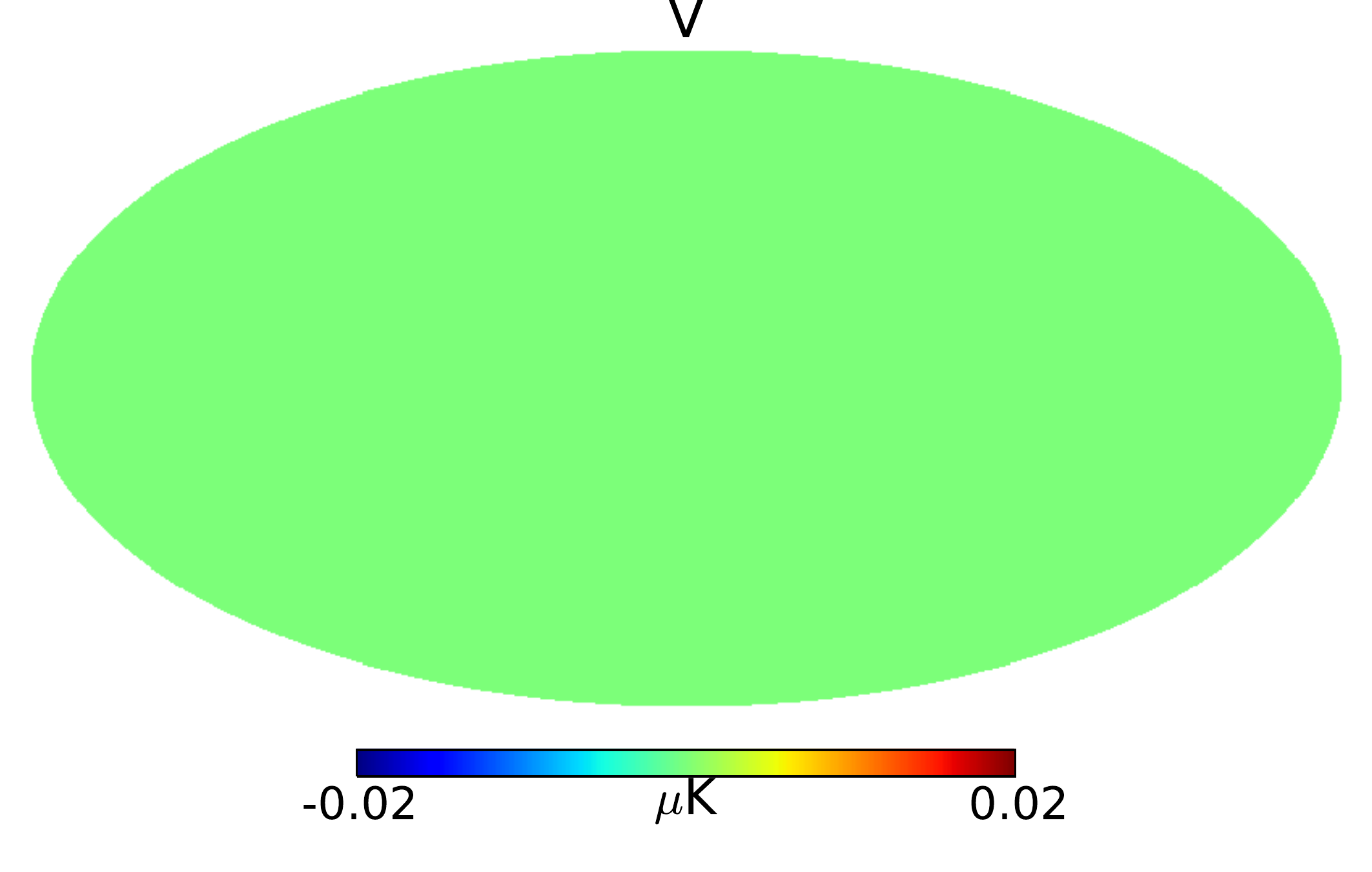} \\
\includegraphics[width=0.33\textwidth]{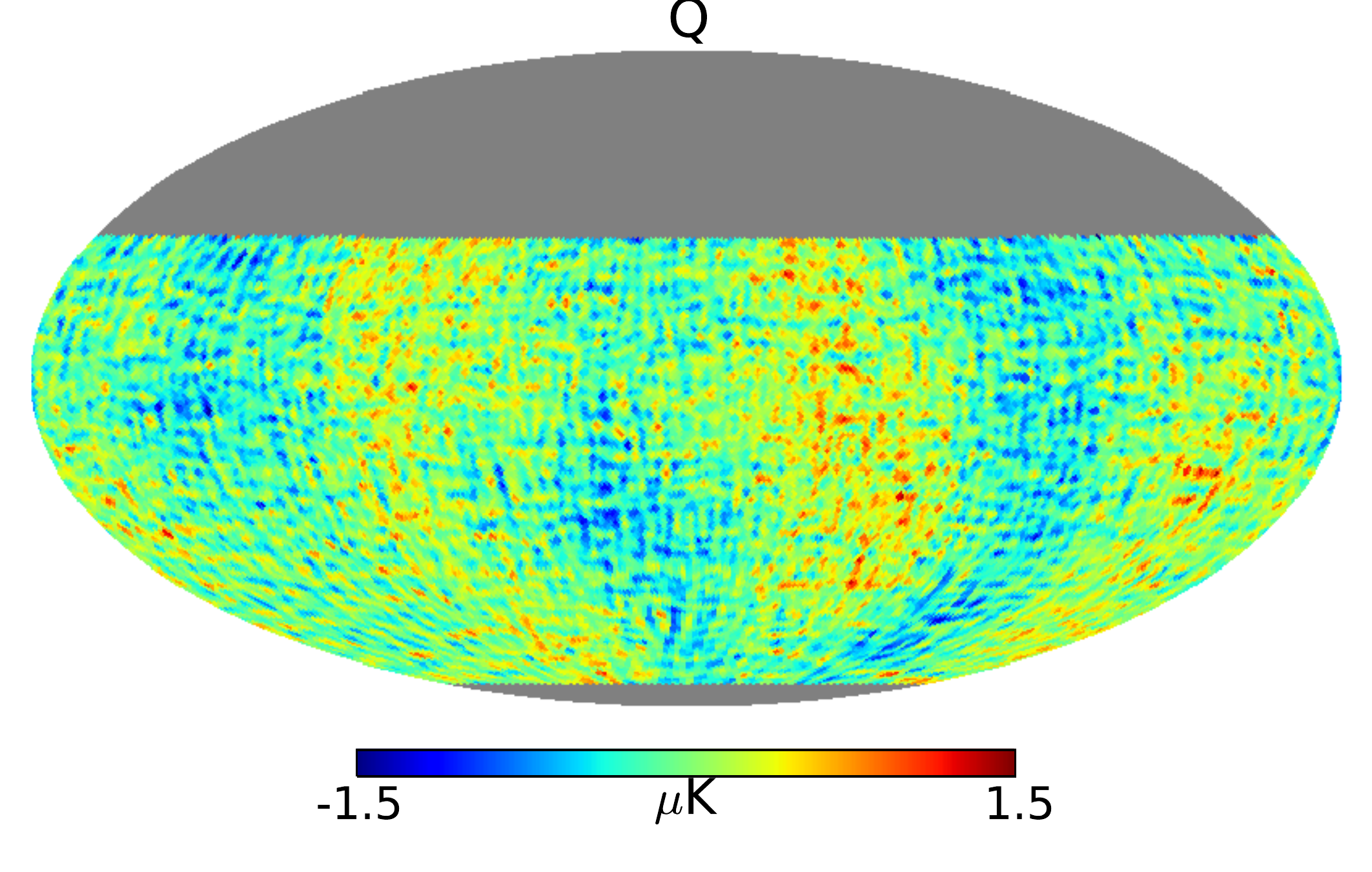} & 
\includegraphics[width=0.33\textwidth]{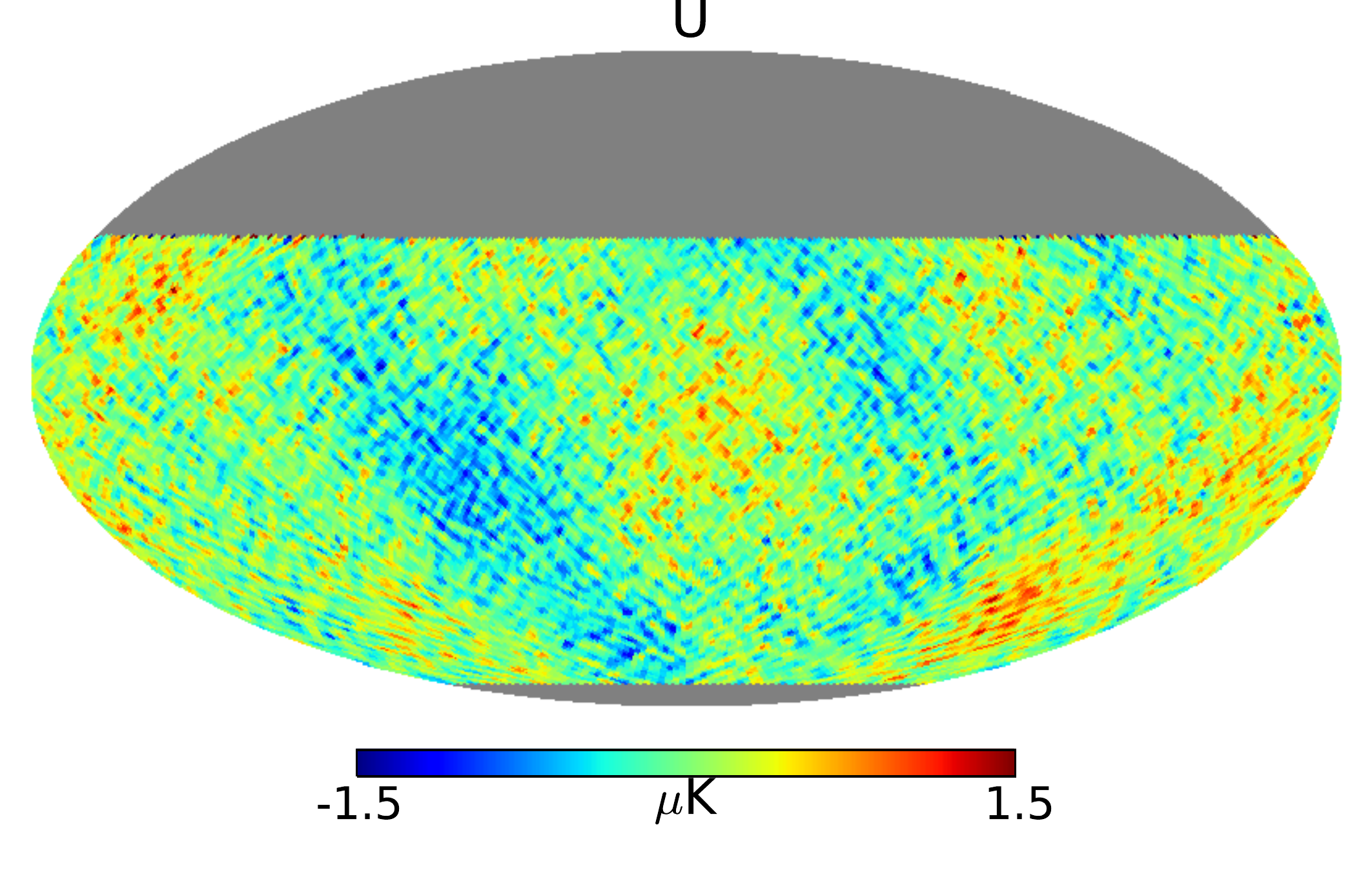} & 
\includegraphics[width=0.33\textwidth]{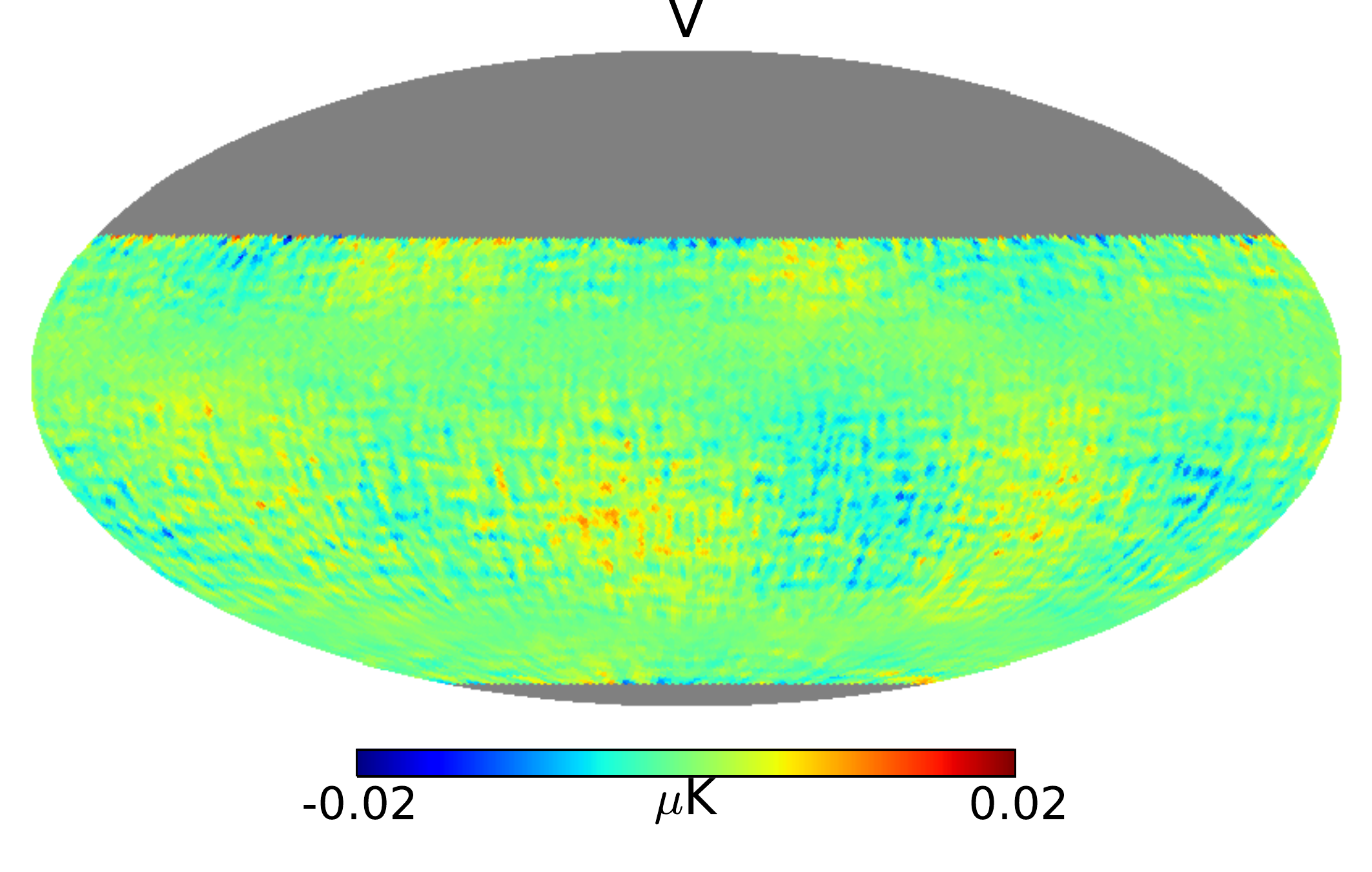} \\
\includegraphics[width=0.33\textwidth]{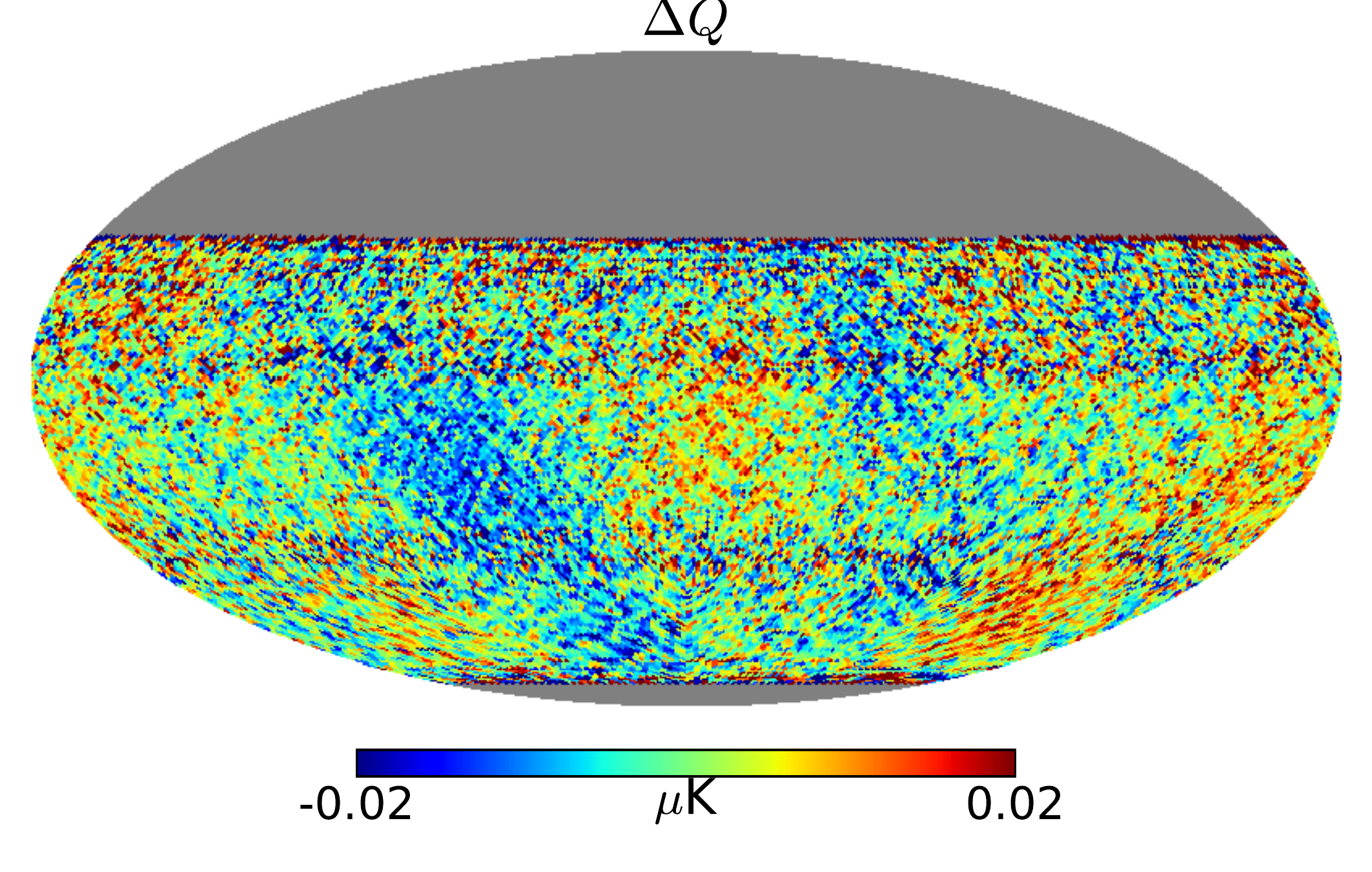} & 
\includegraphics[width=0.33\textwidth]{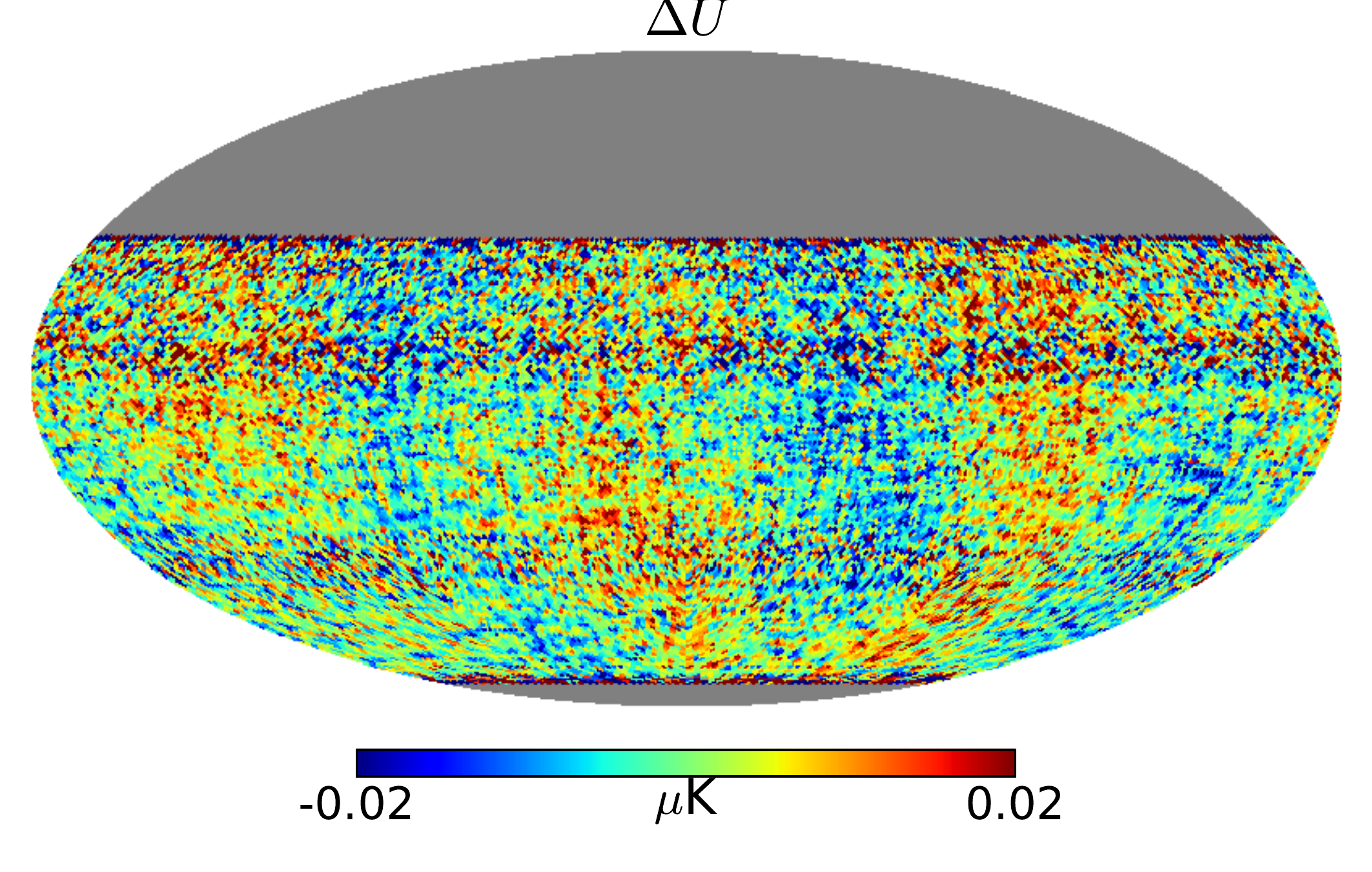} & 
\includegraphics[width=0.33\textwidth]{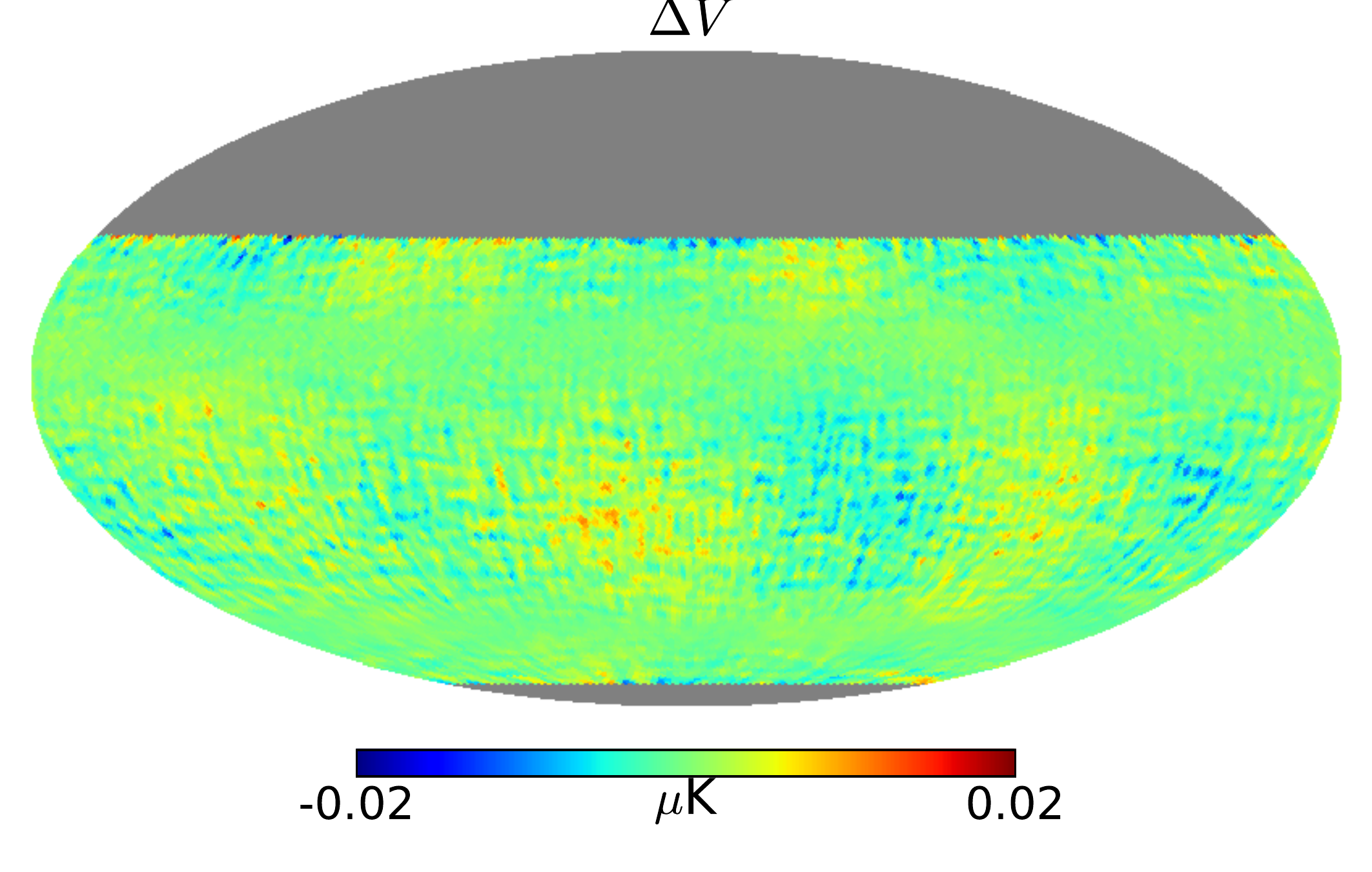}
\end{tabular}
\caption{The top row is the input $Q$, $U$, and $V$ maps for one realization. The middle row is the reconstructed $Q$, $U$, and $V$ maps for 
a single realization of the simulation with the VPM grid at $44.5^{\circ}$ and no VPM emission. The bottom row is the difference between 
the top two rows.}
\label{fig:gridalignmentQUV}
\end{center}
\end{figure*}

Reconstructed $Q$, $U$, and $V$ maps for grid misalignment 
error are shown in \autoref{fig:gridalignmentQUV}. The grid is set to $44.5^{\circ}$ when simulating the data and assumed to be $45^{\circ}$ 
when reconstructing the maps. Here, we do not include the effects of differential emissivity ({\it i.e.}, the VPM physical temperature is set to 
the brightness temperature of the sky in \autoref{eq:gridemission}). An error of $0.5^{\circ}$ is a 
conservative estimate for the uncertainty in grid-detector alignment as this angle will be measurable to greater precision 
than this. There is no visible systematic error dominating the reconstructed $Q$ and $U$ maps; however, the $V$ map shows leakage from 
the input $Q$ map, with the sign depending on the declination. This is due to the slight difference in the $V$ modulation term with the 
grid at $44.5^{\circ}$ and $45.0^{\circ}$. The sign is linked to the scan strategy and how the orientation of the polarization axis of 
the detectors with respect to the sky coordinate system changes over the course of the day.

The amount of $Q$ and $U$ mixing is too small to be evident in just the reconstructed maps; however, some mixing does occur and 
is evident in the difference maps. The $\Delta Q$ map looks like the $U$ map and the $\Delta U$ map looks like $-Q$. Because of 
the misalignment, the measurement of the local $Q$ parameter includes a small contribution from the local $U$ due to the extra 
VPM modulation of $U$ into $Q$. Due to the parallactic angle variation this results in $Q$ in the sky coordinate system having 
some contribution from $U$ and vice-versa. 
In addition, there is a residual coupling into the reconstructed $V$ map
due to small differences between the expected and actual $V$ 
modulation.

\begin{figure}
\begin{center}
\includegraphics[width=0.50\textwidth]{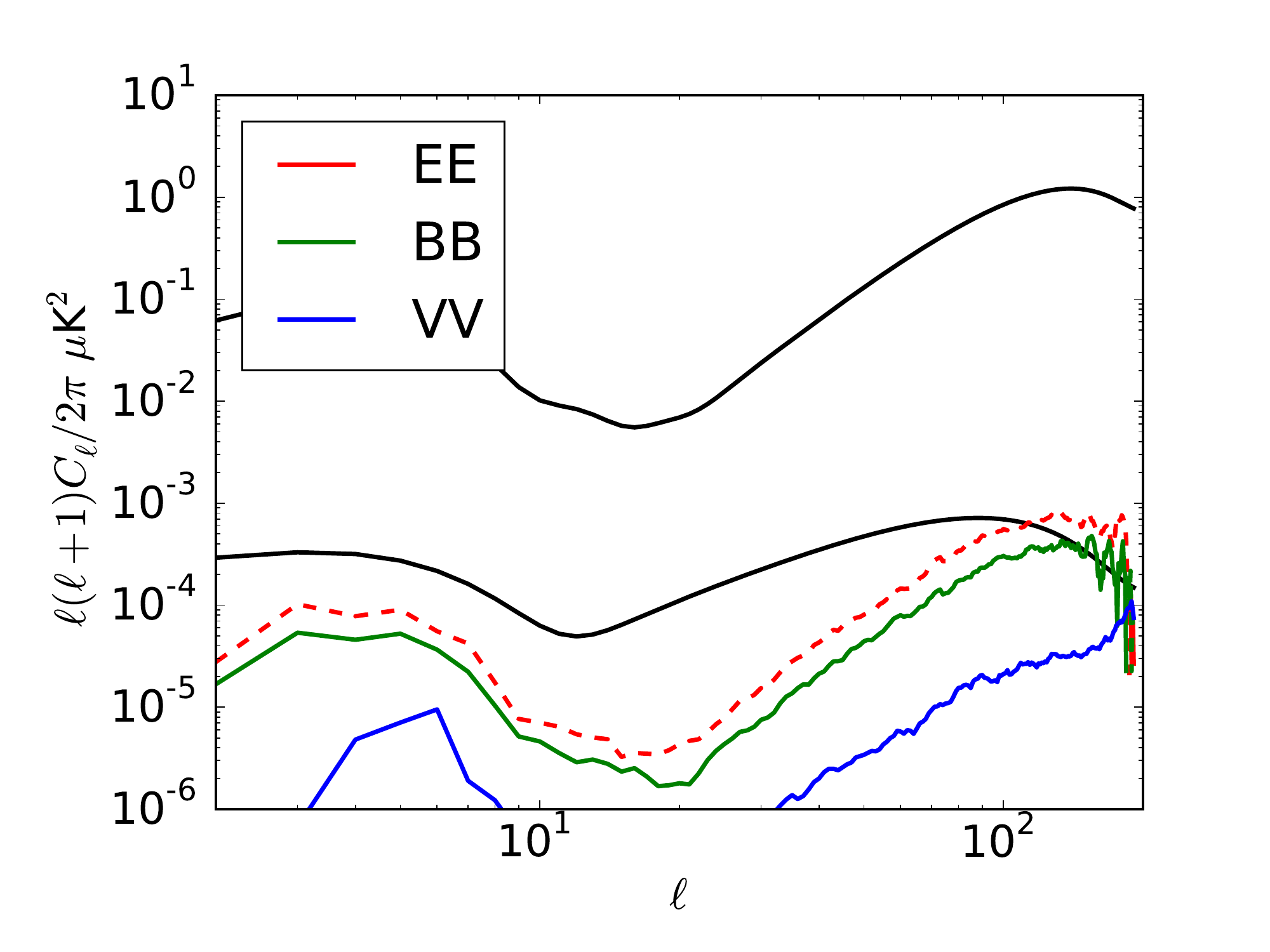}
\caption{Systematic power spectra for the grid misalignment error for a grid-detector angle of $44.5^{\circ}$. 
The red line is the $EE$ systematic power spectrum. The green line is the $BB$ systematic power spectrum. The blue line is the $VV$ systematic power 
spectrum. The solid lines are positive and the dashed lines are negative. 
The black lines are $EE$ (upper) and $BB$ (lower) power spectra for a $\Lambda$CDM model with $r=0.01$. 
The systematic power spectra are 
the difference between the power spectra constructed from simulations with the systematic present and ones without any systematics.
}
\label{fig:gridalignmentps}
\end{center}
\end{figure}

The resulting systematic power spectra are shown in \autoref{fig:gridalignmentps}. The systematic $BB$ power spectrum is $\approx 10\%$ of 
the $B$-mode power for $r=0.01$ on the largest scales. The average systematic over the range $10<\ell<100$ is a few percent of 
$r=0.01$, though it does peak above $r=0.01$ for $100 < \ell < 200$. Both the $BB$ and $VV$ systematic power spectra have shapes similar 
to the $EE$ primordial power spectrum and we can be certain that the $E$-modes are leaking to $B$ and $V$ due to the grid misalignment. 
In addition a small amount of $BB$ would be leaking to $EE$ and $VV$, but its fractional amplitude is small and is not an issue. This 
is expected since the mixing of $Q$ and $U$ in the maps is similar to the case of polarization rotation. The systematic 
$EE$ spectrum is negative because we are comparing the recovered power spectrum to one calculated in a simulation without systematics. 
Since some $E$ leaks to $B$ and $V$, the $EE$ power spectrum exhibits a decrease in amplitude relative to the case in which the grid 
misalignment is absent. 

The $E$ to $B$ leakage introduced by the grid misalignment will generate non-zero $TB$ and $EB$ cross-correlation power spectra. 
This systematic effect can be removed by either implementing a self-calibration technique that employs an \textit{in-situ} measurement 
of the rotation angle using the cross-correlation power spectra \citep{KeatingTBEB2013} or by directly correcting the transfer function 
given a precise measurement of the grid-detector angle. This systematic is due to the fixed misalignment 
between the actual grid-detector angle and the value we used in the transfer functions in the simulation. This single degree of freedom 
can be measured and corrected. For small errors in the grid-detector alignment, the leakage of $E$ to $B$ depends linearly on the size of the misalignment. 
It should be straightforward to correct for the error and lower the amplitude of this systematic below the amplitude of the noise.

\subsection{VPM Emission} \label{s:gridemission}

Our approach for mitigating VPM emission is to remove its signature from the TOD. To do this, it is necessary to understand the functional 
form of the emission as a function of grid-mirror separation. In this section, we describe the generation of the VPM emission template and then show that 
it can be used to clean the TOD in the case of a constant environment. In the next section, we consider VPM emission removal in the presence 
of varying VPM/atmospheric temperature. 

One method of template reconstruction is to lock the detector sampling rate to the VPM throw (e.g. $f_{\rm VPM} = 10 \unit{Hz}$, 
$f_{\rm d} = 100 \unit{Hz}$). In this case, a set number of discrete value of the grid-mirror separation would be repeatedly 
sampled. Because we are using the actual data to extract the VPM emission template, these would be the same positions at which 
the removal of the template would ultimately take place. 

To simulate the effect of VPM emission, we set the angle between the VPM and the detectors to 
be $44.5^{\circ}$, as in the previous section, 
and set the temperature difference between the VPM and the sky to a constant $280 \unit{K}$. 

\begin{figure*}
\begin{center}
\begin{tabular}{ccc}
\includegraphics[width=0.33\textwidth]{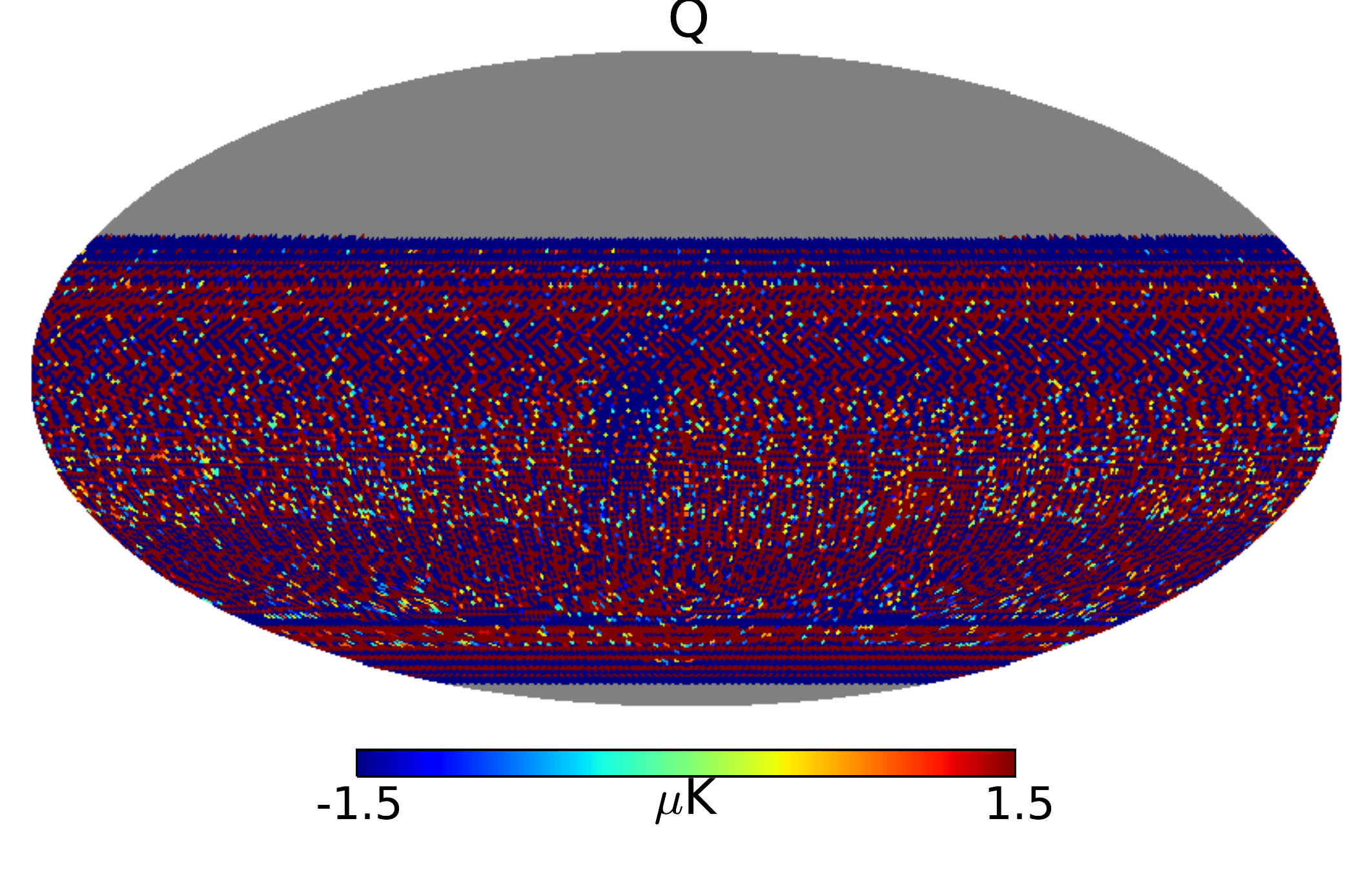} & 
\includegraphics[width=0.33\textwidth]{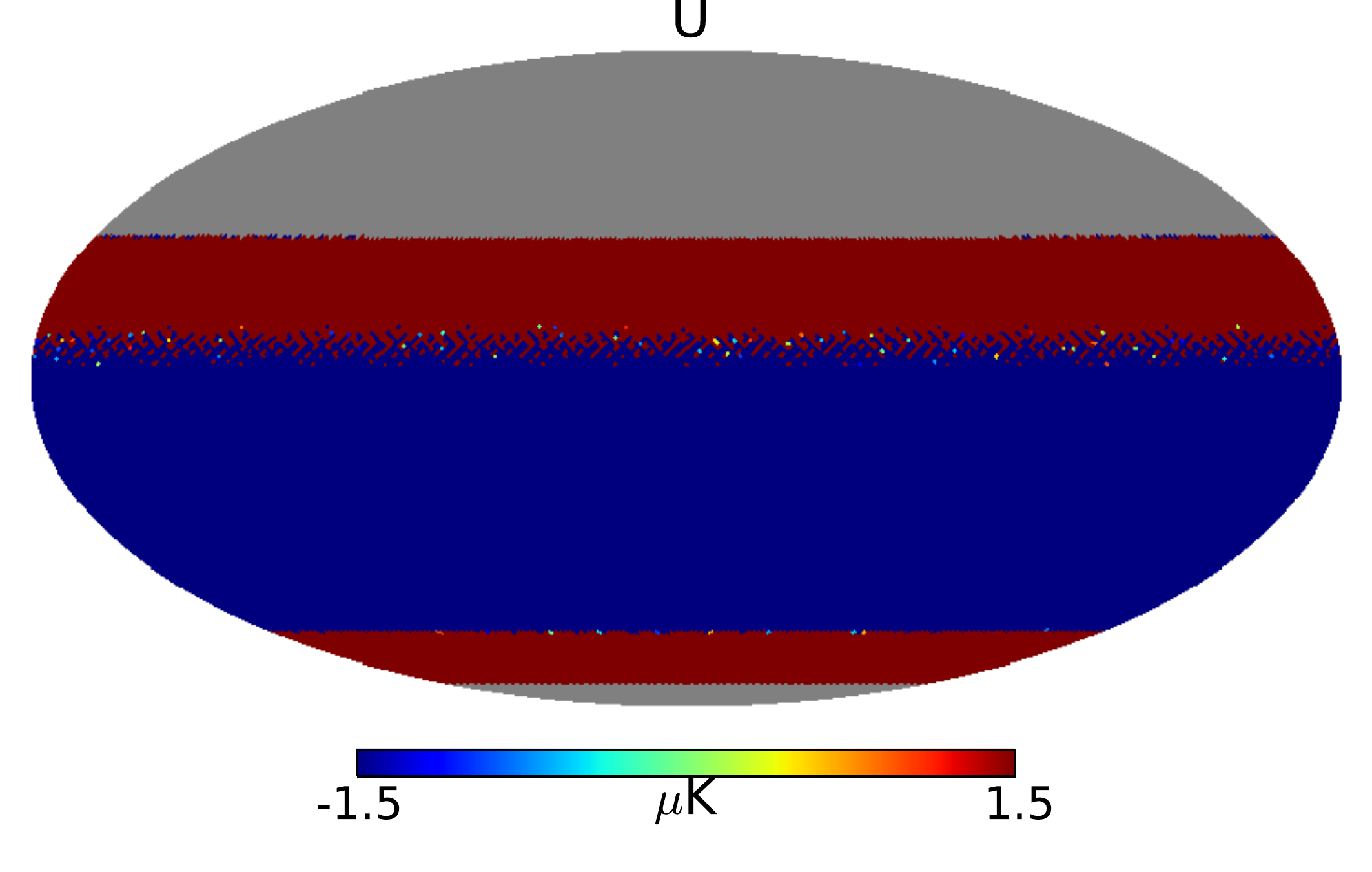} & 
\includegraphics[width=0.33\textwidth]{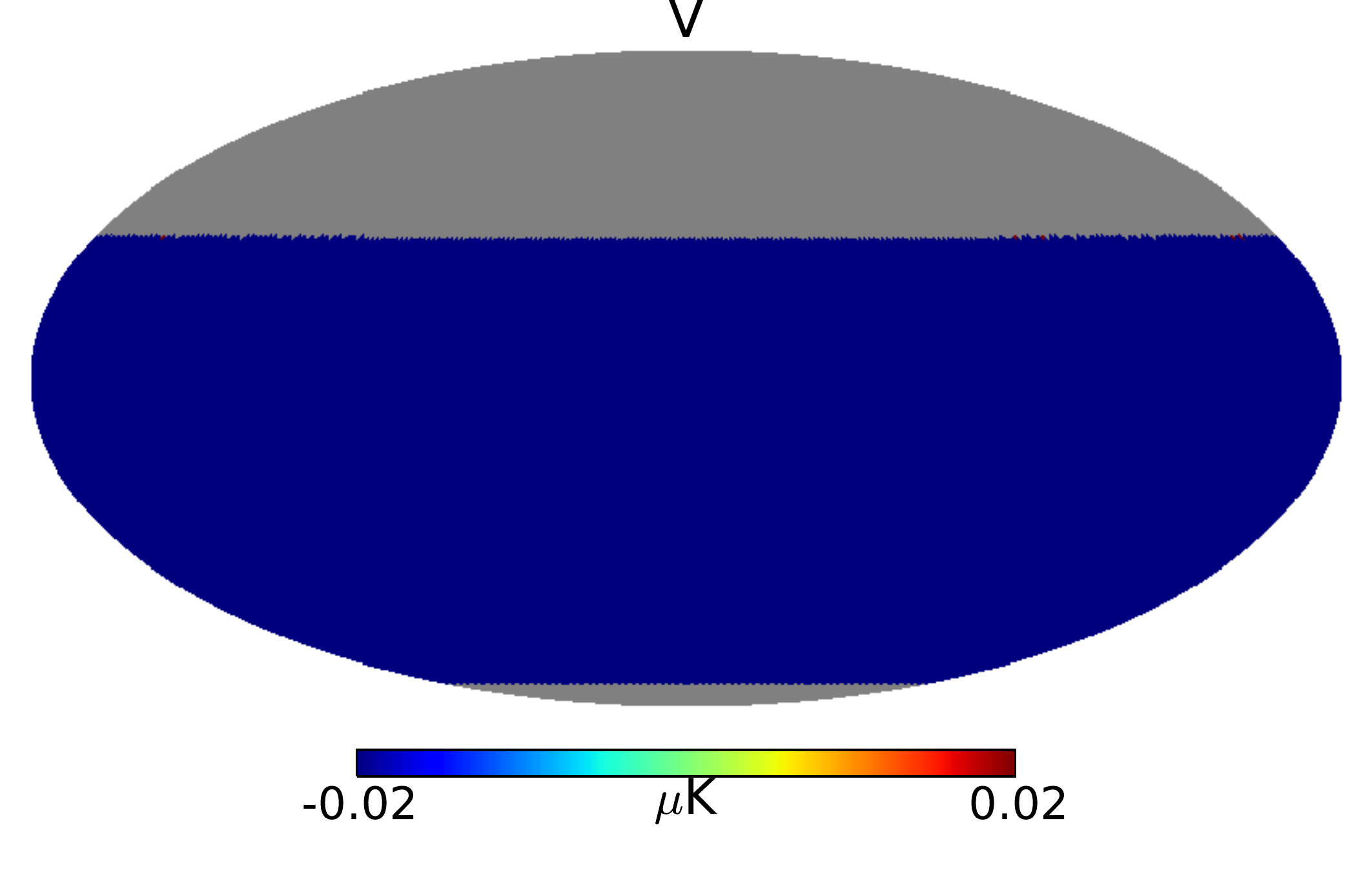} \\
\includegraphics[width=0.33\textwidth]{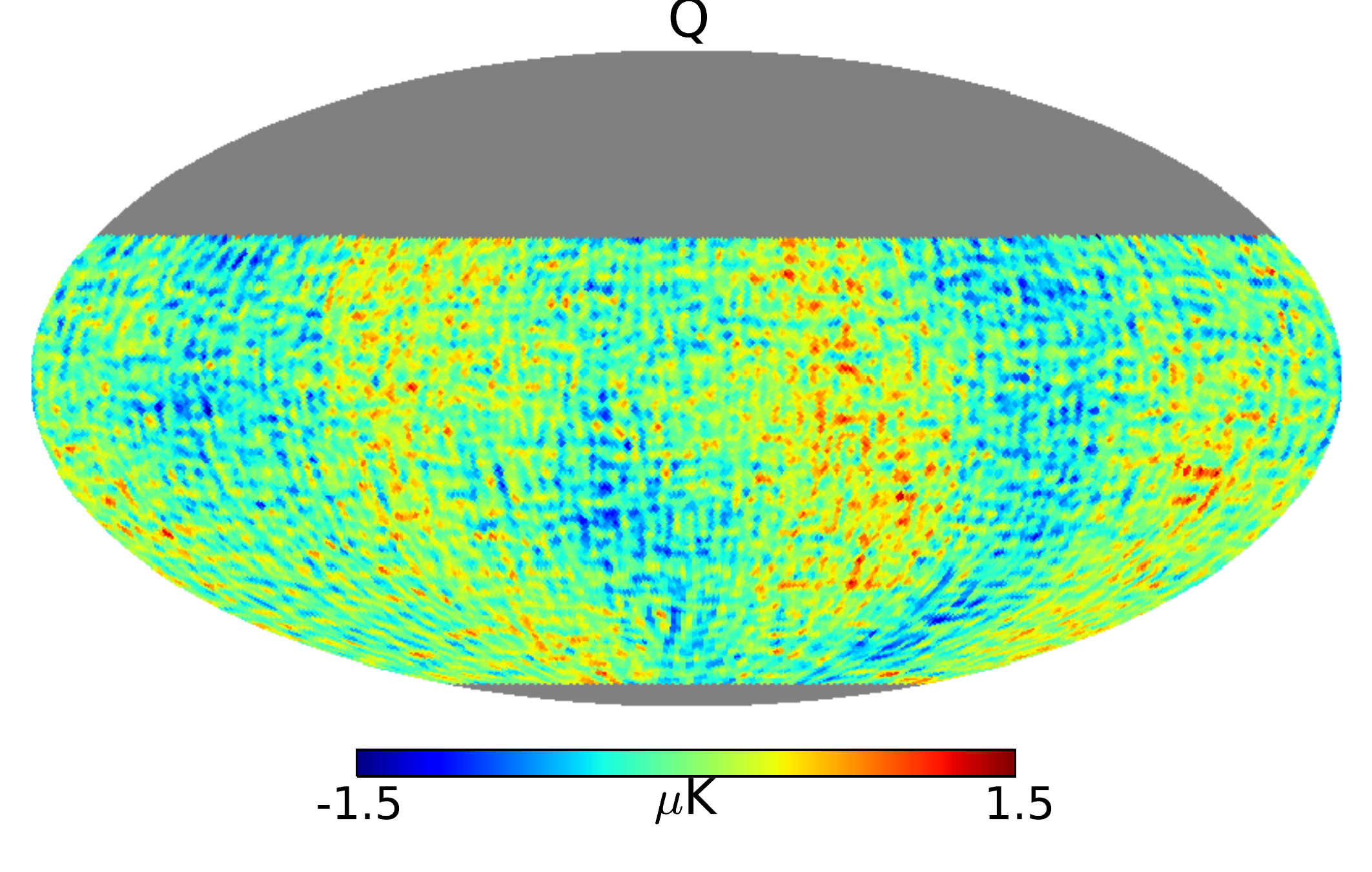} & 
\includegraphics[width=0.33\textwidth]{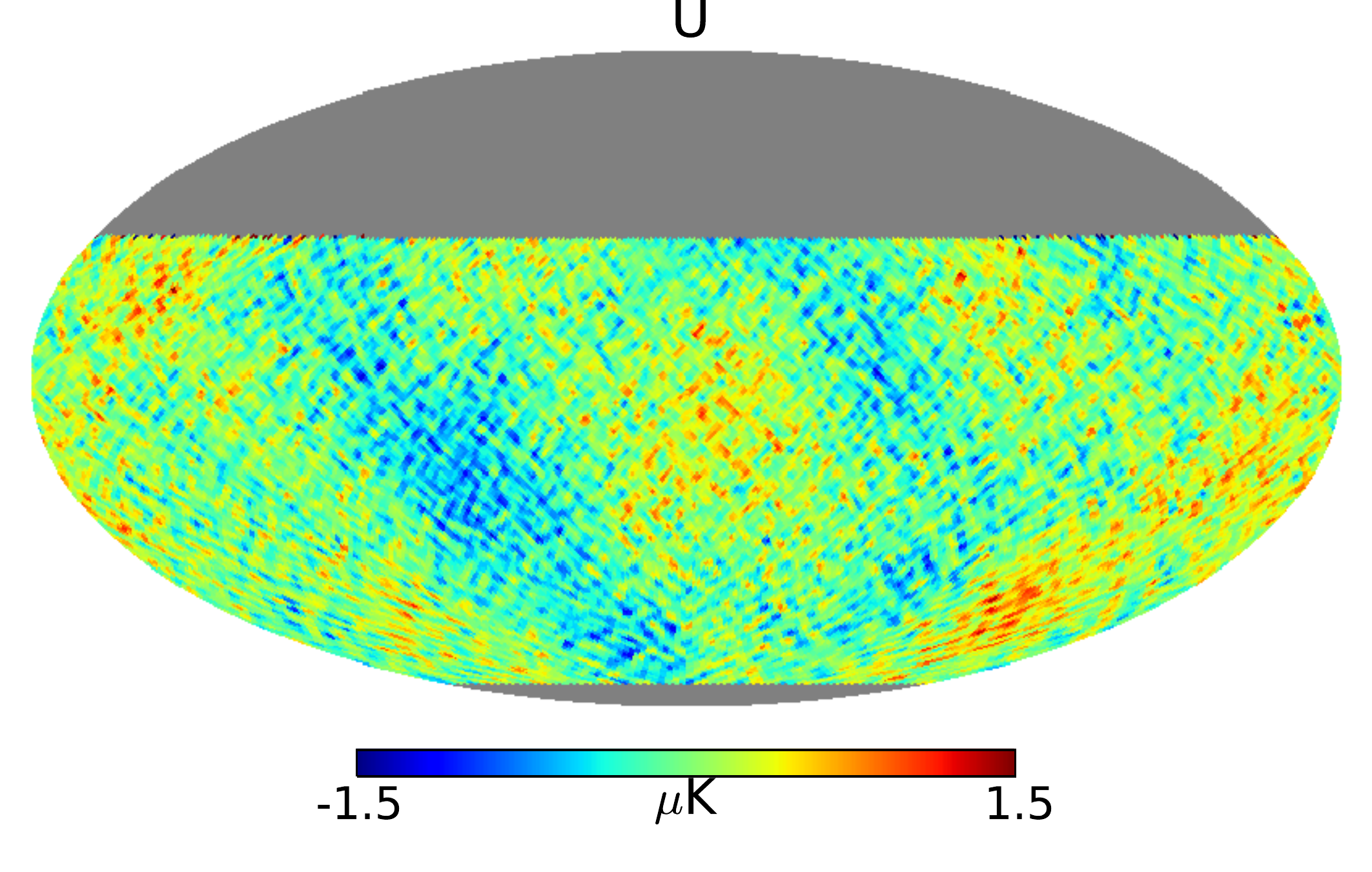} & 
\includegraphics[width=0.33\textwidth]{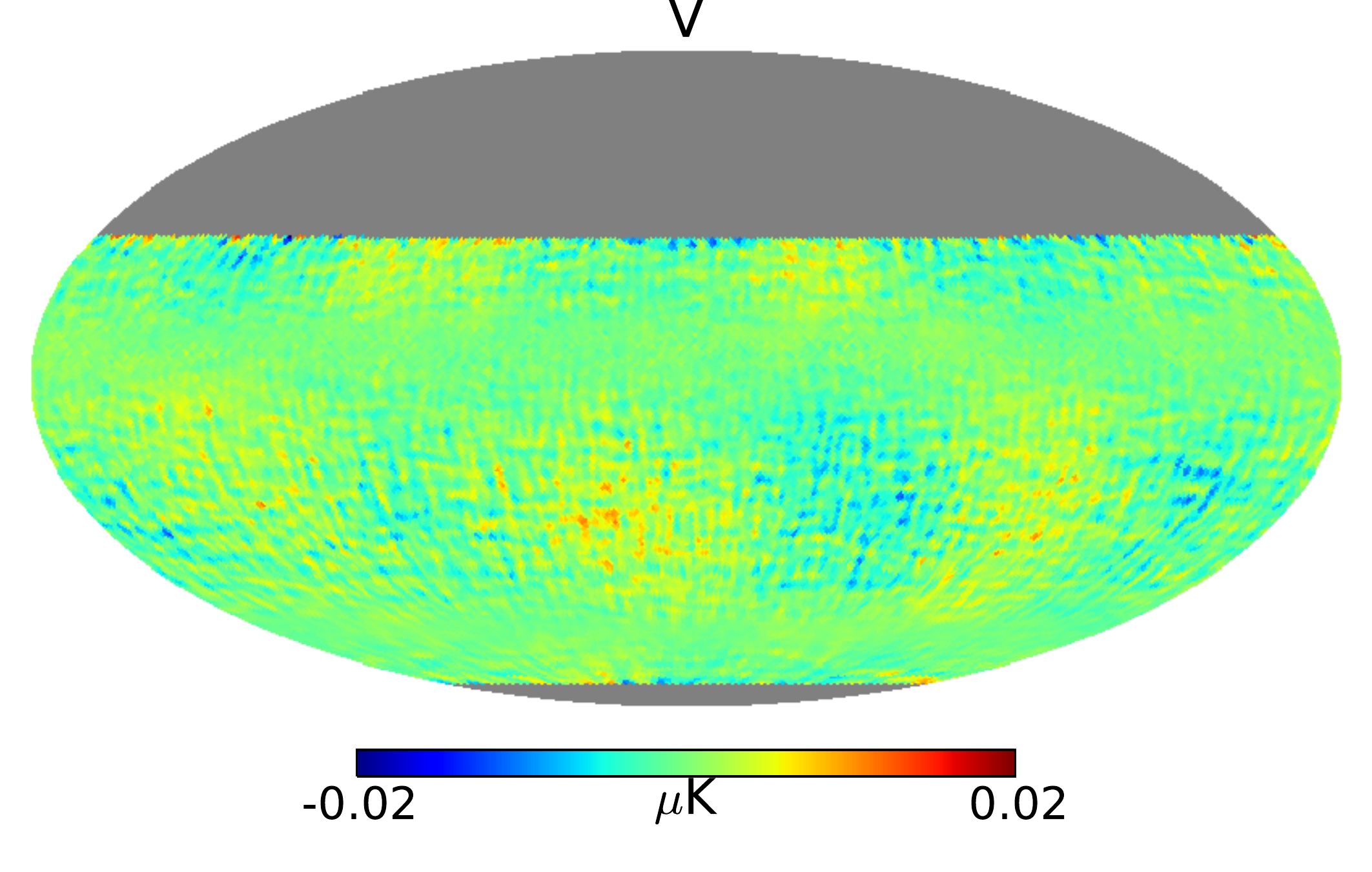}
\end{tabular}
\caption{Reconstructed $Q$, $U$, and $V$ maps for a single realization of the simulation with the VPM grid at $44.5^{\circ}$ and VPM emission from 
a VPM at $280 \unit{K}$. The first row has simulations without VPM emission template subtraction from the TOD. Such simulations are dominated 
by the VPM emission systematic. The second row subtracts the VPM emission residuals from the TOD.}
\label{fig:vpm_emission_tmpl_quv}
\end{center}
\end{figure*}

Reconstructed maps are shown in the top row of \autoref{fig:vpm_emission_tmpl_quv}. To remove the VPM emission from the 
reconstructed maps, we construct a template of the VPM emission using the TOD, and use this template to remove the 
VPM emission from the TOD. We construct the template by partitioning the grid-mirror separation into a number of bins and 
determining the average signal over all detectors for the entire observation in each bin. The sky signal is dependent on the 
position of the detectors and the polarization orientation on the sky, whereas the VPM emission is only a function of grid-mirror separation. 
Therefore, averaging measurements over the entire observation results in the sky signal averaging to zero with the residuals being a good 
estimate of the VPM emission template. This template is then interpolated to the observed grid-mirror positions using a cubic 
interpolation. For grid-mirror positions that are larger than the average value of the last bin or smaller than the average value 
of the first bin and therefore outside the range of interpolation, we apply a quadratic extrapolation. Finally, we subtract the 
interpolated template from the TOD of each detector. 

\begin{figure*}
\begin{center}
\begin{tabular}{cc}
\includegraphics[width=0.50\textwidth]{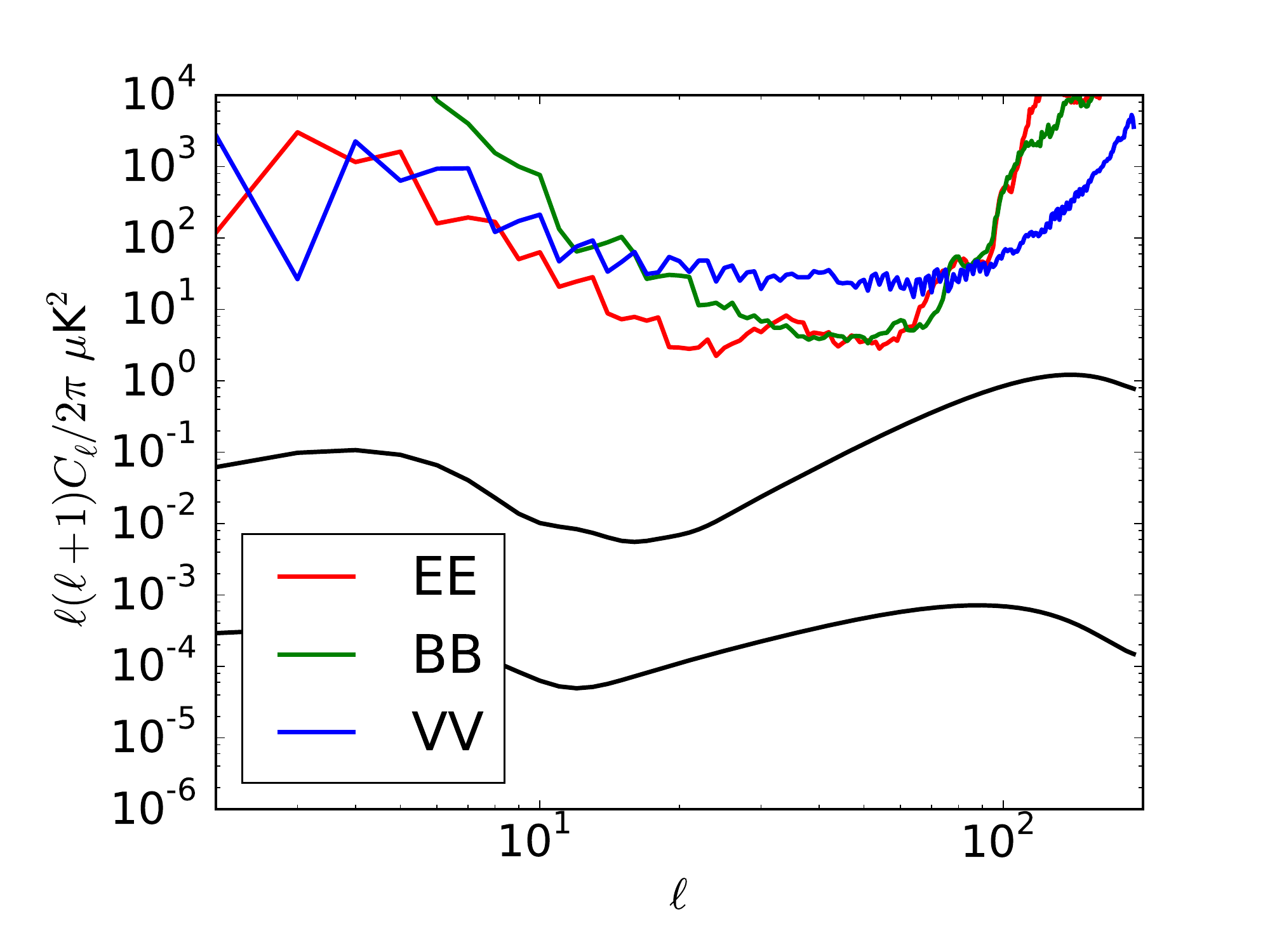} &
\includegraphics[width=0.50\textwidth]{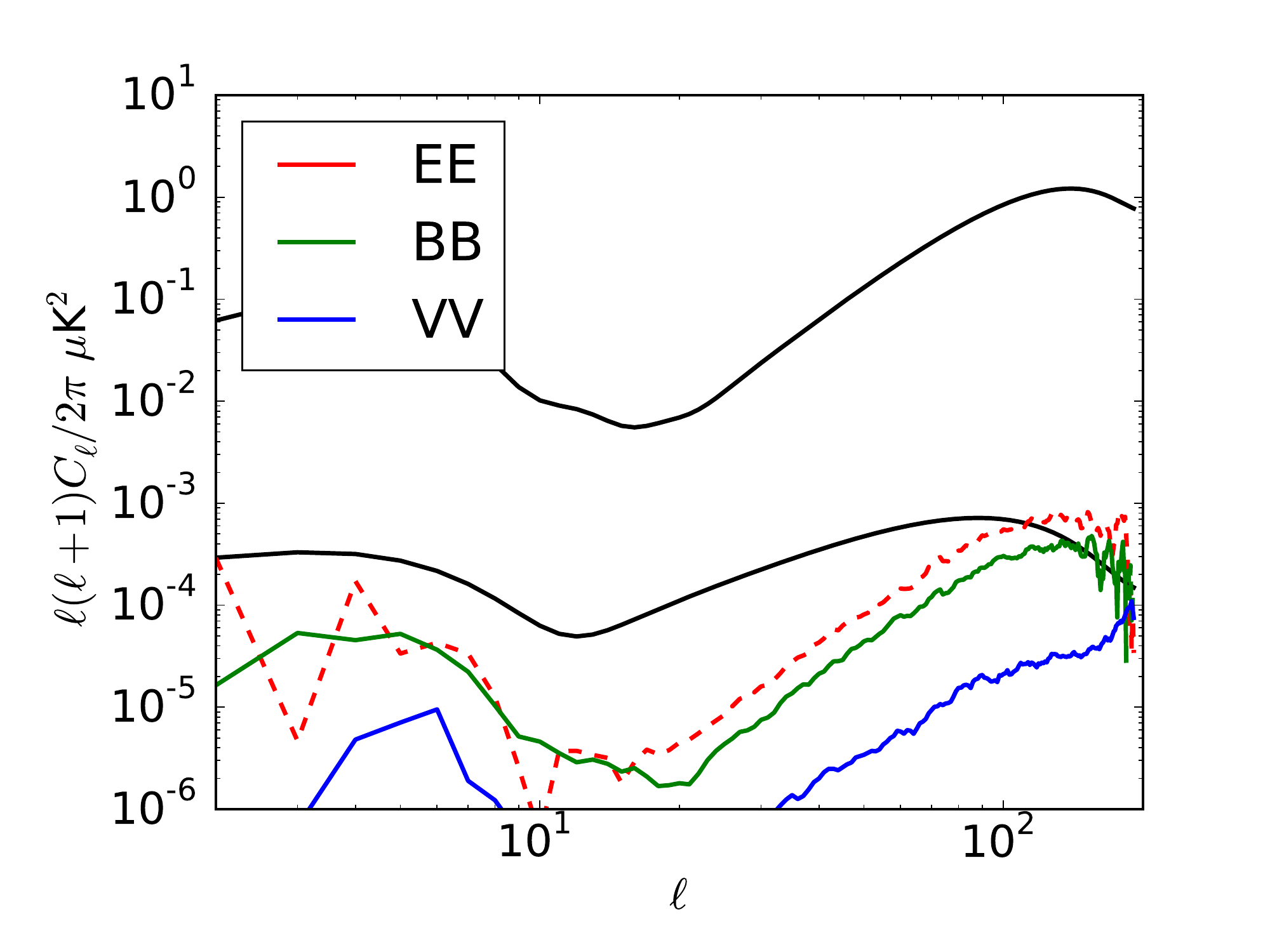} \\
\end{tabular}
\caption{Systematic power spectra for VPM emission with $0.5^{\circ}$ misalignment and a constant VPM temperature. (Left) We do not subtract a template 
from the TOD. (Right) We subtract the VPM emission template from the TOD. In this case, $BB$ systematics 
are dominated by grid misalignment error as in \autoref{fig:gridalignmentps}.
The systematic power spectra are 
the difference between the power spectra constructed from simulations with the systematic present and ones without any systematics.
}
\label{fig:vpm_emission_tmpl_ps}
\end{center}
\end{figure*}

Reconstructed maps with the template subtraction are shown in the bottom row of \autoref{fig:vpm_emission_tmpl_quv}. The template used 
in this simulation has $2000$ bins, $500$ of which contain more than one measurement in the $24$ hour simulation. This number is realizable given the 
$>2600 \unit{\mu m}$ throw and a possible $0.1 \unit{\mu m}$ resolution of the VPM grid-mirror separation. We find that the number of bins can 
be reduced by close to a factor of $10$ while still maintaining the ability to remove the systematic. Plots of the reconstructed power spectra 
in \autoref{fig:vpm_emission_tmpl_ps} demonstrate that the template can be used to successfully reconstruct and remove the VPM emission. The dominant 
remaining systematic in $EE$ and $BB$ power spectra is the leakage due to the grid misalignment. A comparison with \autoref{fig:gridalignmentps} illustrates 
that any residual VPM emission is significantly below the $r=0.01$ level.

Though system (detector and photon) noise for the individual sensors in the instrument is not included in this simulation, it is important to check 
that these factors will not significantly limit this technique. First, the noise introduced by the interpolation can be estimated by the variance 
between the modeled VPM emission and the reconstructed template, which is found to be $\sim 300 \unit{nK}$.  This is significantly below the 
anticipated sensor noise, $200 \unit{\mu K \hspace{1 pt} s^{1/2}}$, in a typical channel. 
Thus, we do not expect the template reconstruction to significantly decrease the sensitivity of the experiment.
Second, this technique assumes that the signal-to-noise is large over time scales short compared to changes in the brightness temperature of the atmosphere.
We anticipate a detection of the $\sim 100 \unit{\mu K}$ VPM emission signal on $\sim$minute time scales. This is significantly below the total 
integration time used to calculate the VPM emission, and thus we anticipate that the presence of noise will not significantly adversely affect the 
result of the template reconstruction described here. This estimate implicitly assumes that the reference phase can be derived from the emission signature 
without incurring additional noise on the time scales of interest. Alternatively, the modulation reference and data acquisition can be synchronized to 
eliminated this potential concern -- in practice this would be the preferred experimental approach.

\subsection{Variable VPM Emission} \label{s:gridemissionTvary}

The amplitude of the VPM emission is directly proportional to the difference between the temperature of the VPM and the 
brightness temperature of the background, which is dominated by the atmosphere. Therefore, the temperature 
excursions in either component will cause the emission signal to vary in time. We generically model these time-dependent 
temperature excursions using a $1/f^2$ power spectrum. The amplitude is chosen to be $2.5 \times 10^{-2} \unit{K/\sqrt{Hz}}$ 
at $1 \unit{Hz}$. This leads to a peak-to-peak variation in the VPM temperature difference of around $2.5 \unit{K}$ when a 
$1.0 \times 10^{-4} \unit{Hz}$ low frequency cutoff set by calibration considerations is applied (i.e. something is 
needed to regularize the spectral powers).  In practice, these variations will be dominated by the atmospheric contribution, and thus this
noise spectrum is effectively modeling fluctuations in the brightness temperature of the atmosphere. The VPM fluctuations are
expected to be much smaller. Components deployed in 
similar thermal environments have observed temperature drifts of $0.5 \unit{K}$ on time scales of hours under 
passive thermal control \citep{Wollack1997}. A real implementation of a VPM should 
include temperature sensor data that can be used in the analysis to understand and remove fluctuations in the VPM emission due to
a varying temperature. Such secondary temperature tracking is not considered in this analysis, as the focus here 
is on the removal of this systematic using radiometric information from the TOD alone. 

With a time-varying temperature difference between the VPM and the sky, the constructed template is the VPM emission at the mean 
temperature. At any point in time, there will be some residual VPM emission whenever the temperature difference is not at its mean 
value. This residual emission may result in an unacceptable level of a systematic $BB$ power spectrum and it is necesary to correct 
for this temperature dependence.

\begin{figure}
\begin{center}
\includegraphics[width=1.0\columnwidth]{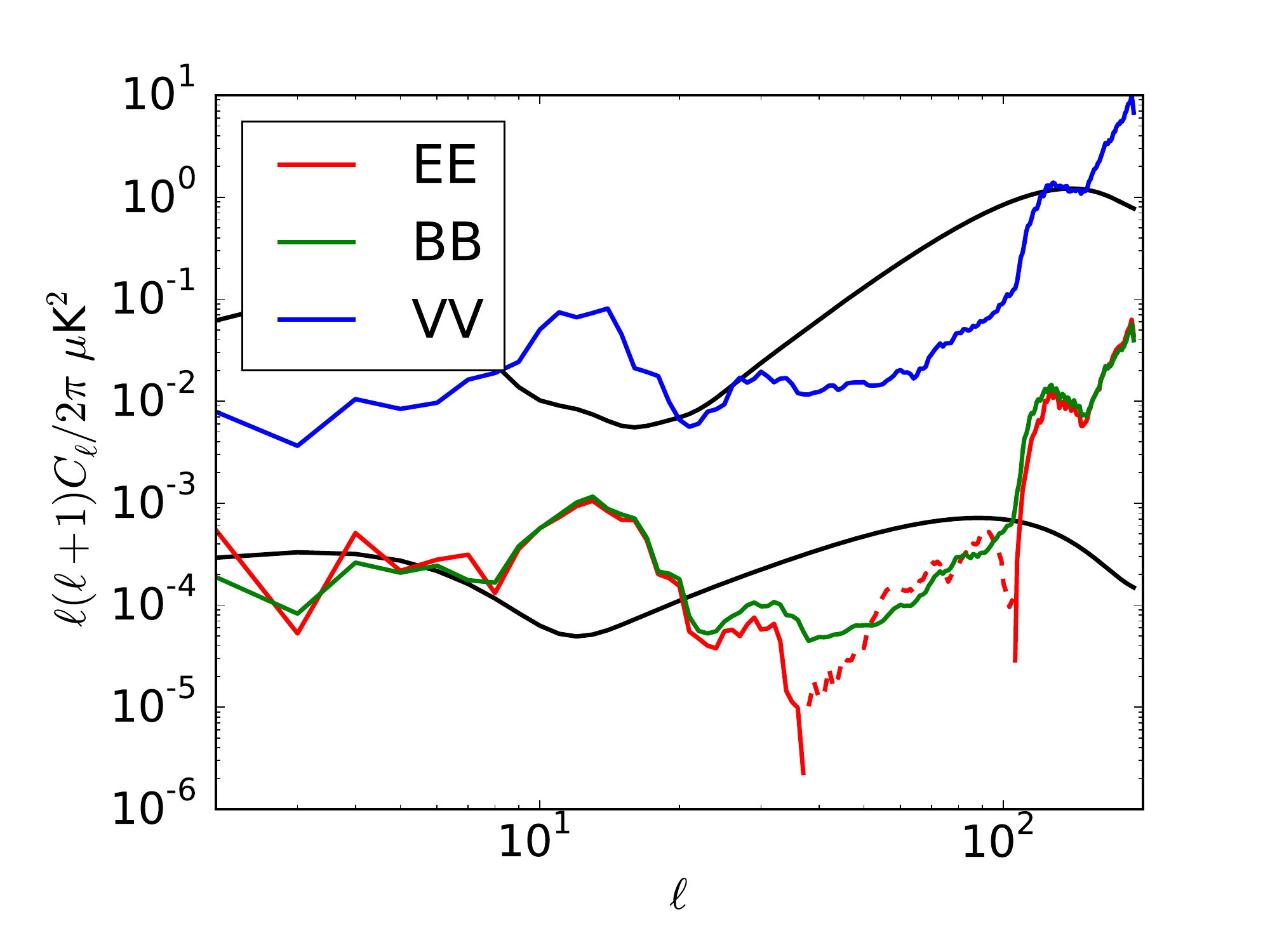}
\caption{Systematic power spectra for a simulation with VPM emission and a varying VPM temperature. 
The systematic power spectra are 
the difference between the power spectra constructed from simulations with the systematic present and ones without any systematics.
An average template as a 
function of VPM grid-mirror separation is subtracted from the TOD to remove the VPM emission. Variation in the VPM emission 
about the average results in systematics comparable to $B$-modes with $r=0.01$. It is possible to mitigate this 
systematic using the destriping algorithm described in the text.
(See \autoref{fig:grid_emission_Tvary2p0_destripe_ps}.)}
\label{fig:grid_emission_Tvary2p0_tmpl_ps}
\end{center}
\end{figure}

For this simulation, we correct for the grid misalignment in the polarization transfer function by using \autoref{eq:vpm_model_1} as 
the model for the TOD. A plot of the reconstructed power spectra with mean teperature template subtraction is shown in 
\autoref{fig:grid_emission_Tvary2p0_tmpl_ps}. The resulting power spectrum is above $r=0.01$ between $\ell = 10$ and $\ell = 20$ 
and would be a concern for measuring $r$ on large angular scales. The amplitude will decrease with additional data since the 
temperature is modeled as a $1/f^2$ noise power spectrum. 

To correct for the residual $1/f^2$ noise from the temperature dependence of the VPM emission, we adopt standard destriping techniques 
\citep{Delabrouille1998,Revenu2000,KurkiSuonio2009,Keihanen2010} to determine the time-dependent temperature difference between the VPM and the sky 
from the TOD. In this case, we have two sets of destriping amplitudes that must be determined for a single length of time. The VPM 
emission as a function of the VPM temperature can be written as a linear combination of a set of amplitudes multiplying the VPM emission at $280 \unit{K}$ 
(template amplitudes) and another set of amplitudes that are offsets to the TOD (offset amplitudes). In other words, the VPM emission at any 
temperature can be written as a function of the VPM emission at $280 \unit{K}$ as 
\begin{equation}
    s_{\rm ve}(T) = a(T) s_{\rm ve}(280 \unit{K}) + b(T)
\end{equation}
where $a$ are the template amplitudes and $b$ are the offsets. Both are functions of the temperature.
The only difference between the template amplitudes and offsets in the absence of other $1/f^2$ noise will be a multiplicative constant as both 
amplitudes should reconstruct the temperature change over time. 
Our implementation of destriping incorporates these two sets of amplitudes. 
For the purposes of this 
simulation, we fit for amplitudes in $20 \unit{s}$ intervals. There are $\approx 200$ full cycles of the VPM over this interval. The sampling 
rates for the various levels of the modulation process are shown in \autoref{table:timescales}. 

\begin{table} \label{table:timescales}
\begin{center}
\begin{tabular}{c|c}
	Sample Rate & Value \\
	\hline Data Sample Rate & $100 \unit{Hz}$ \\
	VPM Sample Rate & $10 \unit{Hz}$ \\
	Beam Scan Rate & $0.5$ - $1.0 \unit{Hz}$ \\
	Destripe Rate & $0.05 \unit{Hz}$ \\
	Calibration Freqeuency & $10^{-4} \unit{Hz}$
\end{tabular}
\caption{The sampling rates for the various levels of the modulation process.}
\end{center}
\end{table}

The full derivation of the destriping procedure is discussed in depth elsewhere \citep{KurkiSuonio2009}. We summarize the equation for the 
optimal amplitudes here. The two sets of amplitudes are the results of solving the equation
\begin{equation} \label{eq:destriping}
	(\mathbf{F}^T \mathbf{N}^{-1} \mathbf{Z} \mathbf{F}) \mathbf{a} = \mathbf{F}^T \mathbf{N}^{-1} \mathbf{Z} \mathbf{d}
\end{equation}
where
\begin{eqnarray}
	\mathbf{Z} &=& \mathbf{I}-\mathbf{A} \mathbf{M}^{-1} \mathbf{A}^T \mathbf{N}^{-1} \nonumber \\
	\mathbf{M} &=& \mathbf{A}^T \mathbf{N}^{-1} \mathbf{A} .
\end{eqnarray}
$\mathbf{A}$, $\mathbf{d}$, and $\mathbf{N}$ were defined for \autoref{eq:map-making}. $\mathbf{I}$ is the identity matrix and 
$\mathbf{F}$ is the matrix that expands the template and offset amplitudes to the TOD. The variable $\mathbf{Z}$ projects out a 
best estimate of the sky signal.
The optimal amplitudes, $\mathbf{a}$, are determined by solving \autoref{eq:destriping} using a preconditioned conjugate gradient method. The best 
fit VPM emission model is then removed from the TOD before estimating the map as in \autoref{eq:map-making}.

\begin{figure}
\begin{center}
\includegraphics[width=1.0\columnwidth]{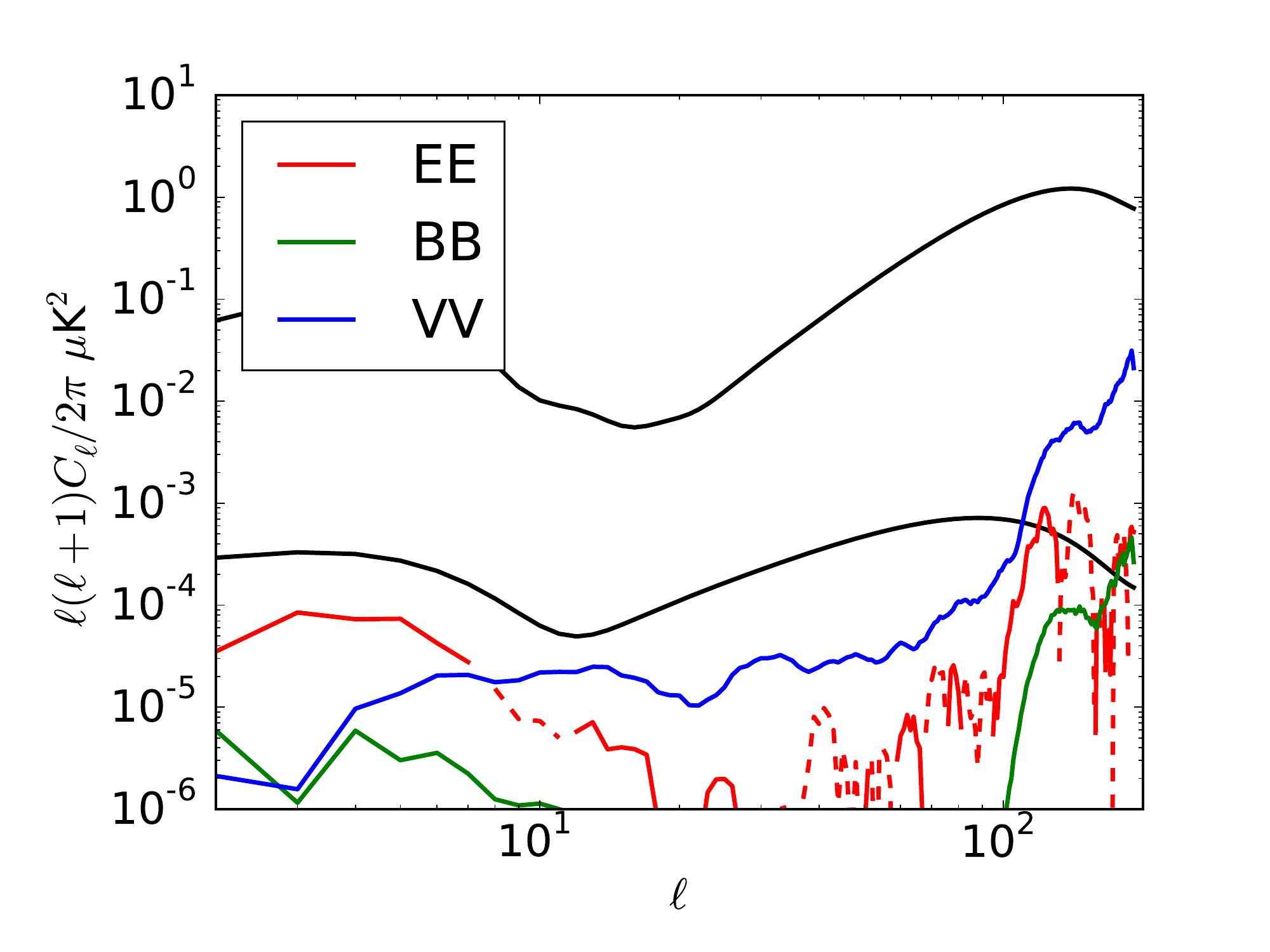}
\caption{Systematic power spectra for a simulation with VPM emission and a varying VPM temperature. 
The systematic power spectra are 
the difference between the power spectra constructed from simulations with the systematic present and ones without any systematics.
The destriping algorithm is used to 
correct the template for the temperature variation to remove the systematic from the TOD. The contamination to $BB$ is seen to be an 
order of magnitude below the $r=0.01$ level.}
\label{fig:grid_emission_Tvary2p0_destripe_ps}
\end{center}
\end{figure}

Power spectra are shown in \autoref{fig:grid_emission_Tvary2p0_destripe_ps} for a simulation in which the TOD are corrected 
using the template subtraction combined with 
the implementation of the destriping algorithm described above. In addition, the polarization transfer function is corrected for 
grid misalignment by including the correct $M_{QQ}$ and $M_{QU}$ Mueller matrix terms in reconstruction. 
In this case, errors in the interpolation of the binned template that couple 
to the time varying temperature necessitate the use of the $2000$ bin template. The 
resulting $BB$ power spectrum is more than an order of magnitude below the target level of $r=0.01$. On the largest 
scales ($\ell < 30$), the residual systematic is on the level of a grid misalignment of $0.1^{\circ}$.

We also consider the case for which the VPM frequency is synchronized to the detector data rate. In this case, we take a VPM frequency of $10 \unit{Hz}$, 
and detector rate of $100 \unit{Hz}$, so that the detector is sampling the same $10$ positions of the VPM throw for every stroke. In this case, 
no interpolation is required and the map residuals are similar to those achieved with the $2000$ bins and an arbitrary sampling scheme.

\section{Conclusions} \label{s:conclusions}

We have run simulations to study how systematics and noise associated with the thermal variation of the atmosphere and VPM will affect measurements 
of the CMB and have highlighted which aspects of a VPM-based experiment need to be measured and/or controlled to reach a desired value of $r$.
We have examined the effect of an error in the VPM transfer function (through an error in the detector-grid alignment), VPM 
emission and absorption, and temperature variation of the VPM and atmosphere on the reconstructed $Q$ and $U$ maps and have propagated these results to the 
$BB$ power spectrum. \textit{None of the systematics studied fundamentally limit the detection of $r$ on large scales for $r=0.01$}. 

The simulations show that a systematic due to grid-angle error may need to be treated for high-fidelity characterization of 
inflationary $B$-modes. An error of $0.5^{\circ}$ in the simulation can be $10\%$ of 
the $B$-mode power at $r=0.01$. Smaller and more realistic angular errors will produce insignificant contamination. 
In the simulations with a time varying VPM temperature, we removed the grid-angle error systematic by 
incorporating a non-zero $M_{\rm QU}$ term into the TOD model, removing most of the grid-angle error systematic. 
Measurements of the VPM transfer function are needed to ensure that this $E$ to $B$ leakage is minimized. Alternatively, the 
grid-angle systematic can be 
determined by fitting for a rotation angle from the $TB$ and $EB$ power spectra and correcting the transfer function similar to the self-calibration 
procedure of \cite{KeatingTBEB2013}. This would preclude a measurement of isotropic cosmic birefringence, so a hardware 
calibration system \citep{Johnson2015} would likely be implemented.

We have shown that the temperature to polarization leakage due to VPM emission can be mitigated by removing a template from the TOD. 
The VPM emission is a function of grid-mirror separation and a high signal-to-noise template can be constructed since we repeatedly observe 
at every different VPM position. 
Different templates may be needed for each separate pixel depending on how the VPM emission changes in detail across the focal plane. 
The template-based emission model remains effective when allowed to vary in time. Implementing destriping with a set of amplitudes that multiply 
the templates allows us to separate most of the time varying systematic from the CMB signal. Additionally, cross-spectra between different 
seasons may alleviate the need for the destriping as the temperature variation would be uncorrelated between the two maps. As long as the 
average template is removed the temperature variations would only add noise to the power spectra instead of extra signal.

While the results shown in this paper assume that the data from the pair of orthogonally-polarized detectors in a pixel are differenced, the 
rapid modulation by the VPM will allow polarization maps to be made from the undifferenced data from a single detector and will allow pixels 
with only one working detector to be used in the analysis. The ability to separate systematics from the signal in this case is currently being 
investigated and preliminary results show that it should not limit the sensitivity of the experiment.

The use of a VPM will enable limits on the Stokes $V$ parameter. This is expected to be a null map in standard $\Lambda$CDM cosmology, but 
there may be some foregrounds \citep{Keating1998,Spinelli2011} that can be measured. The introduction of the $V$ Stokes parameter results 
in four additional power spectra that we can construct: $VV$, $TV$, $EV$, $BV$. We have determined the systematic leakage to the $VV$ power 
spectrum in the simulations. The amplitude of the systematic is on a level similar to that for $EE$ and $BB$. 
For the system and environmental limits considered here, the limit on detecting circularly polarized foregrounds is $\sim 100 \unit{nK}$. 
This is several of orders of magnitude better than the current $95\%$ upper limits on the degree of the CMB circular polarization 
\citep{Mainini2013}.

\section*{Acknowledgements}

Some of the results in this paper have been derived using the HEALPix \citep{Healpix2005} package.

N.J.M.'s research was supported by an appointment to the NASA Postdoctoral Program at Goddard Space Flight Center, administered by Oak Ridge 
Associated Universities through a contract with NASA. K. Harrington was supported by a NASA Space Technology Research Fellowship (NNX14AM49H). 
Support for CLASS has been provided by the National Science Foundation under grant numbers 0959349 
and 1429236.

We thank Al Kogut for helpful comments on a draft of this paper. 

\bibliography{VPM_references}

\begin{thebibliography}{}
\expandafter\ifx\csname natexlab\endcsname\relax\def\natexlab#1{#1}\fi

\bibitem[{{Alexander} {et~al.}(2009){Alexander}, {Ochoa}, \&
  {Kosowsky}}]{Alexander2009}
{Alexander}, S., {Ochoa}, J., \& {Kosowsky}, A. 2009, \prd, 79, 063524

\bibitem[{{Barkats} {et~al.}(2014){Barkats}, {Aikin}, {Bischoff}, {Buder},
  {Kaufman}, {Keating}, {Kovac}, {Su}, {Ade}, {Battle}, {Bierman}, {Bock},
  {Chiang}, {Dowell}, {Duband}, {Filippini}, {Hivon}, {Holzapfel}, {Hristov},
  {Jones}, {Kuo}, {Leitch}, {Mason}, {Matsumura}, {Nguyen}, {Ponthieu},
  {Pryke}, {Richter}, {Rocha}, {Sheehy}, {Kernasovskiy}, {Takahashi}, {Tolan},
  \& {Yoon}}]{BICEP3yr2014}
{Barkats}, D., {Aikin}, R., {Bischoff}, C., {et~al.} 2014, \apj, 783, 67

\bibitem[{{Bennett} {et~al.}(1994){Bennett}, {Kogut}, {Hinshaw}, {Banday},
  {Wright}, {Gorski}, {Wilkinson}, {Weiss}, {Smoot}, {Meyer}, {Mather},
  {Lubin}, {Loewenstein}, {Lineweaver}, {Keegstra}, {Kaita}, {Jackson}, \&
  {Cheng}}]{Bennett1994COBE}
{Bennett}, C.~L., {Kogut}, A., {Hinshaw}, G., {et~al.} 1994, \apj, 436, 423

\bibitem[{{Bennett} {et~al.}(2013){Bennett}, {Larson}, {Weiland}, {Jarosik},
  {Hinshaw}, {Odegard}, {Smith}, {Hill}, {Gold}, {Halpern}, {Komatsu}, {Nolta},
  {Page}, {Spergel}, {Wollack}, {Dunkley}, {Kogut}, {Limon}, {Meyer}, {Tucker},
  \& {Wright}}]{WMAP9yrMaps2013}
{Bennett}, C.~L., {Larson}, D., {Weiland}, J.~L., {et~al.} 2013, \apjs, 208, 20

\bibitem[{{Bersanelli} {et~al.}(2010){Bersanelli}, {Mandolesi}, {Butler},
  {Mennella}, {Villa}, {Aja}, {Artal}, {Artina}, {Baccigalupi}, {Balasini},
  {Baldan}, {Banday}, {Bastia}, {Battaglia}, {Bernardino}, {Blackhurst},
  {Boschini}, {Burigana}, {Cafagna}, {Cappellini}, {Cavaliere}, {Colombo},
  {Crone}, {Cuttaia}, {D'Arcangelo}, {Danese}, {Davies}, {Davis}, {de Angelis},
  {de Gasperis}, {de La Fuente}, {de Rosa}, {de Zotti}, {Falvella}, {Ferrari},
  {Ferretti}, {Figini}, {Fogliani}, {Franceschet}, {Franceschi}, {Gaier},
  {Garavaglia}, {Gomez}, {Gorski}, {Gregorio}, {Guzzi}, {Herreros},
  {Hildebrandt}, {Hoyland}, {Hughes}, {Janssen}, {Jukkala}, {Kettle},
  {Kilpi{\"a}}, {Laaninen}, {Lapolla}, {Lawrence}, {Lawson}, {Leahy},
  {Leonardi}, {Leutenegger}, {Levin}, {Lilje}, {Lowe}, {Lubin}, {Maino},
  {Malaspina}, {Maris}, {Marti-Canales}, {Martinez-Gonzalez}, {Mediavilla},
  {Meinhold}, {Miccolis}, {Morgante}, {Natoli}, {Nesti}, {Pagan}, {Paine},
  {Partridge}, {Pascual}, {Pasian}, {Pearson}, {Pecora}, {Perrotta},
  {Platania}, {Pospieszalski}, {Poutanen}, {Prina}, {Rebolo}, {Roddis},
  {Rubi{\~n}o-Martin}, {Salmon}, {Sandri}, {Seiffert}, {Silvestri},
  {Simonetto}, {Sjoman}, {Smoot}, {Sozzi}, {Stringhetti}, {Taddei}, {Tauber},
  {Terenzi}, {Tomasi}, {Tuovinen}, {Valenziano}, {Varis}, {Vittorio}, {Wade},
  {Wilkinson}, {Winder}, {Zacchei}, \& {Zonca}}]{PlanckLFI2010}
{Bersanelli}, M., {Mandolesi}, N., {Butler}, R.~C., {et~al.} 2010, \aap, 520,
  A4

\bibitem[{{BICEP2 and Keck Array Collaborations} {et~al.}(2015){BICEP2 and Keck
  Array Collaborations}, {:}, {Ade}, {Ahmed}, {Aikin}, {Alexander}, {Barkats},
  {Benton}, {Bischoff}, {Bock}, {Brevik}, {Buder}, {Bullock}, {Buza},
  {Connors}, {Crill}, {Dowell}, {Dvorkin}, {Duband}, {Filippini}, {Fliescher},
  {Golwala}, {Halpern}, {Harrison}, {Hasselfield}, {Hildebrandt}, {Hilton},
  {Hristov}, {Hui}, {Irwin}, {Karkare}, {Kaufman}, {Keating}, {Kefeli},
  {Kernasovskiy}, {Kovac}, {Kuo}, {Leitch}, {Lueker}, {Mason}, {Megerian},
  {Netterfield}, {Nguyen}, {O'Brient}, {Ogburn}, {Orlando}, {Pryke},
  {Reintsema}, {Richter}, {Schwarz}, {Sheehy}, {Staniszewski}, {Sudiwala},
  {Teply}, {Thompson}, {Tolan}, {Turner}, {Vieregg}, {Weber}, {Willmert},
  {Wong}, \& {Yoon}}]{Keck2015}
{BICEP2 and Keck Array Collaborations}, {:}, {Ade}, P.~A.~R., {et~al.} 2015,
  \apj, 811, 126

\bibitem[{{BICEP2 and Keck Array Collaborations} {et~al.}(2016){BICEP2 and Keck
  Array Collaborations}, {:}, {Ade}, {Ahmed}, {Aikin}, {Alexander}, {Barkats},
  {Benton}, {Bischoff}, {Bock}, {Bowens-Rubin}, {Brevik}, {Buder}, {Bullock},
  {Buza}, {Connors}, {Crill}, {Duband}, {Dvorkin}, {Filippini}, {Fliescher},
  {Grayson}, {Halpern}, {Harrison}, {Hilton}, {Hui}, {Irwin}, {Karkare},
  {Karpel}, {Kaufman}, {Keating}, {Kefeli}, {Kernasovskiy}, {Kovac}, {Kuo},
  {Leitch}, {Lueker}, {Megerian}, {Netterfield}, {Nguyen}, {O'Brient},
  {Ogburn}, {Orlando}, {Pryke}, {Richter}, {Schwarz}, {Sheehy}, {Staniszewski},
  {Steinbach}, {Sudiwala}, {Teply}, {Thompson}, {Tolan}, {Tucker}, {Turner},
  {Vieregg}, {Weber}, {Wiebe}, {Willmert}, {Wong}, {Wu}, \&
  {Yoon}}]{BICEP2KeckVI2015}
---. 2016, Physical Review Letters, 116, 031302

\bibitem[{{BICEP2 Collaboration} {et~al.}(2014{\natexlab{a}}){BICEP2
  Collaboration}, {Ade}, {Aikin}, {Amiri}, {Barkats}, {Benton}, {Bischoff},
  {Bock}, {Brevik}, {Buder}, {Bullock}, {Davis}, {Day}, {Dowell}, {Duband},
  {Filippini}, {Fliescher}, {Golwala}, {Halpern}, {Hasselfield}, {Hildebrandt},
  {Hilton}, {Irwin}, {Karkare}, {Kaufman}, {Keating}, {Kernasovskiy}, {Kovac},
  {Kuo}, {Leitch}, {Llombart}, {Lueker}, {Netterfield}, {Nguyen}, {O'Brient},
  {Ogburn}, {Orlando}, {Pryke}, {Reintsema}, {Richter}, {Schwarz}, {Sheehy},
  {Staniszewski}, {Story}, {Sudiwala}, {Teply}, {Tolan}, {Turner}, {Vieregg},
  {Wilson}, {Wong}, {Yoon}, \& {Bicep2 Collaboration}}]{BICEP2II2014}
{BICEP2 Collaboration}, {Ade}, P.~A.~R., {Aikin}, R.~W., {et~al.}
  2014{\natexlab{a}}, \apj, 792, 62

\bibitem[{{BICEP2 Collaboration} {et~al.}(2014{\natexlab{b}}){BICEP2
  Collaboration}, {Ade}, {Aikin}, {Barkats}, {Benton}, {Bischoff}, {Bock},
  {Brevik}, {Buder}, {Bullock}, {Dowell}, {Duband}, {Filippini}, {Fliescher},
  {Golwala}, {Halpern}, {Hasselfield}, {Hildebrandt}, {Hilton}, {Hristov},
  {Irwin}, {Karkare}, {Kaufman}, {Keating}, {Kernasovskiy}, {Kovac}, {Kuo},
  {Leitch}, {Lueker}, {Mason}, {Netterfield}, {Nguyen}, {O'Brient}, {Ogburn},
  {Orlando}, {Pryke}, {Reintsema}, {Richter}, {Schwarz}, {Sheehy},
  {Staniszewski}, {Sudiwala}, {Teply}, {Tolan}, {Turner}, {Vieregg}, {Wong}, \&
  {Yoon}}]{BICEP22014}
---. 2014{\natexlab{b}}, Physical Review Letters, 112, 241101

\bibitem[{{Chuss} {et~al.}(2012){Chuss}, {Wollack}, {Henry}, {Hui}, {Juarez},
  {Krejny}, {Moseley}, \& {Novak}}]{Chuss2012}
{Chuss}, D.~T., {Wollack}, E.~J., {Henry}, R., {et~al.} 2012, \ao, 51, 197

\bibitem[{{Chuss} {et~al.}(2006){Chuss}, {Wollack}, {Moseley}, \&
  {Novak}}]{ChussVPM2006}
{Chuss}, D.~T., {Wollack}, E.~J., {Moseley}, S.~H., \& {Novak}, G. 2006, \ao,
  45, 5107

\bibitem[{{Chuss} {et~al.}(2010){Chuss}, {Ade}, {Benford}, {Bennett}, {Dotson},
  {Eimer}, {Fixsen}, {Halpern}, {Hilton}, {Hinderks}, {Hinshaw}, {Irwin},
  {Jackson}, {Jah}, {Jethava}, {Jhabvala}, {Kogut}, {Lowe}, {McCullagh},
  {Miller}, {Mirel}, {Moseley}, {Rodriguez}, {Rostem}, {Sharp}, {Staguhn},
  {Tucker}, {Voellmer}, {Wollack}, \& {Zeng}}]{Chuss2010SPIE}
{Chuss}, D.~T., {Ade}, P.~A.~R., {Benford}, D.~J., {et~al.} 2010, in \procspie,
  Vol. 7741, Millimeter, Submillimeter, and Far-Infrared Detectors and
  Instrumentation for Astronomy V, 77411P

\bibitem[{{Cooray} {et~al.}(2003){Cooray}, {Melchiorri}, \&
  {Silk}}]{Cooray2003}
{Cooray}, A., {Melchiorri}, A., \& {Silk}, J. 2003, Physics Letters B, 554, 1

\bibitem[{{Delabrouille}(1998)}]{Delabrouille1998}
{Delabrouille}, J. 1998, \aaps, 127, 555

\bibitem[{Dicke(1946)}]{Dicke1946}
Dicke, R.~H. 1946, Review of Scientific Instruments, 17, 268

\bibitem[{{D{\"u}nner} {et~al.}(2013){D{\"u}nner}, {Hasselfield}, {Marriage},
  {Sievers}, {Acquaviva}, {Addison}, {Ade}, {Aguirre}, {Amiri}, {Appel},
  {Barrientos}, {Battistelli}, {Bond}, {Brown}, {Burger}, {Calabrese},
  {Chervenak}, {Das}, {Devlin}, {Dicker}, {Bertrand Doriese}, {Dunkley},
  {Essinger-Hileman}, {Fisher}, {Gralla}, {Fowler}, {Hajian}, {Halpern},
  {Hern{\'a}ndez-Monteagudo}, {Hilton}, {Hilton}, {Hincks}, {Hlozek},
  {Huffenberger}, {Hughes}, {Hughes}, {Infante}, {Irwin}, {Baptiste Juin},
  {Kaul}, {Klein}, {Kosowsky}, {Lau}, {Limon}, {Lin}, {Louis}, {Lupton},
  {Marsden}, {Martocci}, {Mauskopf}, {Menanteau}, {Moodley}, {Moseley},
  {Netterfield}, {Niemack}, {Nolta}, {Page}, {Parker}, {Partridge}, {Quintana},
  {Reid}, {Sehgal}, {Sherwin}, {Spergel}, {Staggs}, {Swetz}, {Switzer},
  {Thornton}, {Trac}, {Tucker}, {Warne}, {Wilson}, {Wollack}, \&
  {Zhao}}]{DunnerACT2013}
{D{\"u}nner}, R., {Hasselfield}, M., {Marriage}, T.~A., {et~al.} 2013, \apj,
  762, 10

\bibitem[{{Eimer} {et~al.}(2012){Eimer}, {Bennett}, {Chuss}, {Marriage},
  {Wollack}, \& {Zeng}}]{CLASS2012SPIE}
{Eimer}, J.~R., {Bennett}, C.~L., {Chuss}, D.~T., {et~al.} 2012, in \procspie,
  Vol. 8452, Millimeter, Submillimeter, and Far-Infrared Detectors and
  Instrumentation for Astronomy VI, 845220

\bibitem[{{Emerson} \& {Graeve}(1988)}]{Emerson1988}
{Emerson}, D.~T., \& {Graeve}, R. 1988, \aap, 190, 353

\bibitem[{{Emerson} {et~al.}(1979){Emerson}, {Klein}, \&
  {Haslam}}]{Emerson1979}
{Emerson}, D.~T., {Klein}, U., \& {Haslam}, C.~G.~T. 1979, \aap, 76, 92

\bibitem[{{Errard} {et~al.}(2015){Errard}, {Ade}, {Akiba}, {Arnold}, {Atlas},
  {Baccigalupi}, {Barron}, {Boettger}, {Borrill}, {Chapman}, {Chinone},
  {Cukierman}, {Delabrouille}, {Dobbs}, {Ducout}, {Elleflot}, {Fabbian},
  {Feng}, {Feeney}, {Gilbert}, {Goeckner-Wald}, {Halverson}, {Hasegawa},
  {Hattori}, {Hazumi}, {Hill}, {Holzapfel}, {Hori}, {Inoue}, {Jaehnig},
  {Jaffe}, {Jeong}, {Katayama}, {Kaufman}, {Keating}, {Kermish}, {Keskitalo},
  {Kisner}, {Le Jeune}, {Lee}, {Leitch}, {Leon}, {Linder}, {Matsuda},
  {Matsumura}, {Miller}, {Myers}, {Navaroli}, {Nishino}, {Okamura}, {Paar},
  {Peloton}, {Poletti}, {Puglisi}, {Rebeiz}, {Reichardt}, {Richards}, {Ross},
  {Rotermund}, {Schenck}, {Sherwin}, {Siritanasak}, {Smecher}, {Stebor},
  {Steinbach}, {Stompor}, {Suzuki}, {Tajima}, {Takakura}, {Tikhomirov},
  {Tomaru}, {Whitehorn}, {Wilson}, {Yadav}, \& {Zahn}}]{Errard2015}
{Errard}, J., {Ade}, P.~A.~R., {Akiba}, Y., {et~al.} 2015, \apj, 809, 63

\bibitem[{{Essinger-Hileman} {et~al.}(2014){Essinger-Hileman}, {Ali}, {Amiri},
  {Appel}, {Araujo}, {Bennett}, {Boone}, {Chan}, {Cho}, {Chuss}, {Colazo},
  {Crowe}, {Denis}, {D{\"u}nner}, {Eimer}, {Gothe}, {Halpern}, {Harrington},
  {Hilton}, {Hinshaw}, {Huang}, {Irwin}, {Jones}, {Karakla}, {Kogut}, {Larson},
  {Limon}, {Lowry}, {Marriage}, {Mehrle}, {Miller}, {Miller}, {Moseley},
  {Novak}, {Reintsema}, {Rostem}, {Stevenson}, {Towner}, {U-Yen}, {Wagner},
  {Watts}, {Wollack}, {Xu}, \& {Zeng}}]{CLASS2014SPIE}
{Essinger-Hileman}, T., {Ali}, A., {Amiri}, M., {et~al.} 2014, in \procspie,
  Vol. 9153, Millimeter, Submillimeter, and Far-Infrared Detectors and
  Instrumentation for Astronomy VII, 91531I

\bibitem[{{Fraisse} {et~al.}(2013){Fraisse}, {Ade}, {Amiri}, {Benton}, {Bock},
  {Bond}, {Bonetti}, {Bryan}, {Burger}, {Chiang}, {Clark}, {Contaldi}, {Crill},
  {Davis}, {Dor{\'e}}, {Farhang}, {Filippini}, {Fissel}, {Gandilo}, {Golwala},
  {Gudmundsson}, {Hasselfield}, {Hilton}, {Holmes}, {Hristov}, {Irwin},
  {Jones}, {Kuo}, {MacTavish}, {Mason}, {Montroy}, {Morford}, {Netterfield},
  {O'Dea}, {Rahlin}, {Reintsema}, {Ruhl}, {Runyan}, {Schenker}, {Shariff},
  {Soler}, {Trangsrud}, {Tucker}, {Tucker}, {Turner}, \&
  {Wiebe}}]{Spider2013JCAP}
{Fraisse}, A.~A., {Ade}, P.~A.~R., {Amiri}, M., {et~al.} 2013, \jcap, 4, 47

\bibitem[{{G{\'o}rski} {et~al.}(2005){G{\'o}rski}, {Hivon}, {Banday},
  {Wandelt}, {Hansen}, {Reinecke}, \& {Bartelmann}}]{Healpix2005}
{G{\'o}rski}, K.~M., {Hivon}, E., {Banday}, A.~J., {et~al.} 2005, \apj, 622,
  759

\bibitem[{{Grain} {et~al.}(2009){Grain}, {Tristram}, \& {Stompor}}]{Grain2009}
{Grain}, J., {Tristram}, M., \& {Stompor}, R. 2009, \prd, 79, 123515

\bibitem[{{Hanany} \& {Rosenkranz}(2003)}]{HananyRosenkranz2003}
{Hanany}, S., \& {Rosenkranz}, P. 2003, \nar, 47, 1159

\bibitem[{{Hinderks} {et~al.}(2009){Hinderks}, {Ade}, {Bock}, {Bowden},
  {Brown}, {Cahill}, {Carlstrom}, {Castro}, {Church}, {Culverhouse},
  {Friedman}, {Ganga}, {Gear}, {Gupta}, {Harris}, {Haynes}, {Keating}, {Kovac},
  {Kirby}, {Lange}, {Leitch}, {Mallie}, {Melhuish}, {Memari}, {Murphy},
  {Orlando}, {Schwarz}, {Sullivan}, {Piccirillo}, {Pryke}, {Rajguru},
  {Rusholme}, {Taylor}, {Thompson}, {Tucker}, {Turner}, {Wu}, \&
  {Zemcov}}]{HinderksQUaD2009}
{Hinderks}, J.~R., {Ade}, P., {Bock}, J., {et~al.} 2009, \apj, 692, 1221

\bibitem[{{Hinshaw} {et~al.}(2013){Hinshaw}, {Larson}, {Komatsu}, {Spergel},
  {Bennett}, {Dunkley}, {Nolta}, {Halpern}, {Hill}, {Odegard}, {Page}, {Smith},
  {Weiland}, {Gold}, {Jarosik}, {Kogut}, {Limon}, {Meyer}, {Tucker}, {Wollack},
  \& {Wright}}]{WMAP9yrCosmo2013}
{Hinshaw}, G., {Larson}, D., {Komatsu}, E., {et~al.} 2013, \apjs, 208, 19

\bibitem[{{Hu} \& {Okamoto}(2002)}]{HuOkamoto2002}
{Hu}, W., \& {Okamoto}, T. 2002, \apj, 574, 566

\bibitem[{{Jarosik} {et~al.}(2003){Jarosik}, {Bennett}, {Halpern}, {Hinshaw},
  {Kogut}, {Limon}, {Meyer}, {Page}, {Pospieszalski}, {Spergel}, {Tucker},
  {Wilkinson}, {Wollack}, {Wright}, \& {Zhang}}]{WMAPRadiometers2003}
{Jarosik}, N., {Bennett}, C.~L., {Halpern}, M., {et~al.} 2003, \apjs, 145, 413

\bibitem[{{Johnson} {et~al.}(2015){Johnson}, {Vourch}, {Drysdale}, {Kalman},
  {Fujikawa}, {Keating}, \& {Kaufman}}]{Johnson2015}
{Johnson}, B.~R., {Vourch}, C.~J., {Drysdale}, T.~D., {et~al.} 2015, Journal of
  Astronomical Instrumentation, 4, 50007

\bibitem[{{Kamionkowski} {et~al.}(1997){Kamionkowski}, {Kosowsky}, \&
  {Stebbins}}]{Kamionkowski1997b}
{Kamionkowski}, M., {Kosowsky}, A., \& {Stebbins}, A. 1997, \prd, 55, 7368

\bibitem[{{Karakci} {et~al.}(2013){Karakci}, {Zhang}, {Sutter}, {Bunn},
  {Korotkov}, {Timbie}, {Tucker}, \& {Wandelt}}]{Karakci2013}
{Karakci}, A., {Zhang}, L., {Sutter}, P.~M., {et~al.} 2013, \apjs, 207, 14

\bibitem[{{Keating} {et~al.}(1998){Keating}, {Timbie}, {Polnarev}, \&
  {Steinberger}}]{Keating1998}
{Keating}, B., {Timbie}, P., {Polnarev}, A., \& {Steinberger}, J. 1998, \apj,
  495, 580

\bibitem[{{Keating} {et~al.}(2003){Keating}, {O'Dell}, {Gundersen},
  {Piccirillo}, {Stebor}, \& {Timbie}}]{POLAR2003}
{Keating}, B.~G., {O'Dell}, C.~W., {Gundersen}, J.~O., {et~al.} 2003, \apjs,
  144, 1

\bibitem[{{Keating} {et~al.}(2013){Keating}, {Shimon}, \&
  {Yadav}}]{KeatingTBEB2013}
{Keating}, B.~G., {Shimon}, M., \& {Yadav}, A.~P.~S. 2013, \apjl, 762, L23

\bibitem[{{Keih{\"a}nen} {et~al.}(2010){Keih{\"a}nen}, {Keskitalo},
  {Kurki-Suonio}, {Poutanen}, \& {Sirvi{\"o}}}]{Keihanen2010}
{Keih{\"a}nen}, E., {Keskitalo}, R., {Kurki-Suonio}, H., {Poutanen}, T., \&
  {Sirvi{\"o}}, A.-S. 2010, \aap, 510, A57

\bibitem[{{Kosowsky}(1996)}]{Kosowsky1996}
{Kosowsky}, A. 1996, Annals of Physics, 246, 49

\bibitem[{{Kurki-Suonio} {et~al.}(2009){Kurki-Suonio}, {Keih{\"a}nen},
  {Keskitalo}, {Poutanen}, {Sirvi{\"o}}, {Maino}, \&
  {Burigana}}]{KurkiSuonio2009}
{Kurki-Suonio}, H., {Keih{\"a}nen}, E., {Keskitalo}, R., {et~al.} 2009, \aap,
  506, 1511

\bibitem[{{Kusaka} {et~al.}(2014){Kusaka}, {Essinger-Hileman}, {Appel},
  {Gallardo}, {Irwin}, {Jarosik}, {Nolta}, {Page}, {Parker}, {Raghunathan},
  {Sievers}, {Simon}, {Staggs}, \& {Visnjic}}]{ABS_HWP_2013}
{Kusaka}, A., {Essinger-Hileman}, T., {Appel}, J.~W., {et~al.} 2014, Review of
  Scientific Instruments, 85, 024501

\bibitem[{{Lazear} {et~al.}(2014){Lazear}, {Ade}, {Benford}, {Bennett},
  {Chuss}, {Dotson}, {Eimer}, {Fixsen}, {Halpern}, {Hilton}, {Hinderks},
  {Hinshaw}, {Irwin}, {Jhabvala}, {Johnson}, {Kogut}, {Lowe}, {McMahon},
  {Miller}, {Mirel}, {Moseley}, {Rodriguez}, {Sharp}, {Staguhn}, {Switzer},
  {Tucker}, {Weston}, \& {Wollack}}]{PIPER2014SPIE}
{Lazear}, J., {Ade}, P.~A.~R., {Benford}, D., {et~al.} 2014, in \procspie, Vol.
  9153, Millimeter, Submillimeter, and Far-Infrared Detectors and
  Instrumentation for Astronomy VII, 91531L

\bibitem[{{MacTavish} {et~al.}(2008){MacTavish}, {Ade}, {Battistelli},
  {Benton}, {Bihary}, {Bock}, {Bond}, {Brevik}, {Bryan}, {Contaldi}, {Crill},
  {Dor{\'e}}, {Fissel}, {Golwala}, {Halpern}, {Hilton}, {Holmes}, {Hristov},
  {Irwin}, {Jones}, {Kuo}, {Lange}, {Lawrie}, {Martin}, {Mason}, {Montroy},
  {Netterfield}, {Riley}, {Ruhl}, {Runyan}, {Trangsrud}, {Tucker}, {Turner},
  {Viero}, \& {Wiebe}}]{MacTavish2008}
{MacTavish}, C.~J., {Ade}, P.~A.~R., {Battistelli}, E.~S., {et~al.} 2008, \apj,
  689, 655

\bibitem[{{Mainini} {et~al.}(2013){Mainini}, {Minelli}, {Gervasi}, {Boella},
  {Sironi}, {Ba{\'u}}, {Banfi}, {Passerini}, {De Lucia}, \&
  {Cavaliere}}]{Mainini2013}
{Mainini}, R., {Minelli}, D., {Gervasi}, M., {et~al.} 2013, \jcap, 8, 33

\bibitem[{{Mangum} {et~al.}(2007){Mangum}, {Emerson}, \&
  {Greisen}}]{Magnum2007}
{Mangum}, J.~G., {Emerson}, D.~T., \& {Greisen}, E.~W. 2007, \aap, 474, 679

\bibitem[{{Miller} {et~al.}(2009){Miller}, {Shimon}, \& {Keating}}]{Miller2009}
{Miller}, N.~J., {Shimon}, M., \& {Keating}, B.~G. 2009, \prd, 79, 063008

\bibitem[{{Moyerman} {et~al.}(2013){Moyerman}, {Bierman}, {Ade}, {Aiken},
  {Barkats}, {Bischoff}, {Bock}, {Chiang}, {Dowell}, {Duband}, {Hivon},
  {Holzapfel}, {Hristov}, {Jones}, {Kaufman}, {Keating}, {Kovac}, {Kuo},
  {Leitch}, {Mason}, {Matsumura}, {Nguyen}, {Ponthieu}, {Pryke}, {Richter},
  {Rocha}, {Sheehy}, {Takahashi}, {Tolan}, {Wollack}, \& {Yoon}}]{Moyerman2013}
{Moyerman}, S., {Bierman}, E., {Ade}, P.~A.~R., {et~al.} 2013, \apj, 765, 64

\bibitem[{{Planck Collaboration} {et~al.}(2015){Planck Collaboration}, {Ade},
  {Aghanim}, {Arnaud}, {Ashdown}, {Aumont}, {Baccigalupi}, {Banday},
  {Barreiro}, {Bartlett}, \& et~al.}]{Planck2015Cosmo}
{Planck Collaboration}, {Ade}, P.~A.~R., {Aghanim}, N., {et~al.} 2015, ArXiv
  e-prints, arXiv:1502.01589

\bibitem[{{POLARBEAR Collaboration} {et~al.}(2014){POLARBEAR Collaboration},
  {Ade}, {Akiba}, {Anthony}, {Arnold}, {Atlas}, {Barron}, {Boettger},
  {Borrill}, {Chapman}, {Chinone}, {Dobbs}, {Elleflot}, {Errard}, {Fabbian},
  {Feng}, {Flanigan}, {Gilbert}, {Grainger}, {Halverson}, {Hasegawa},
  {Hattori}, {Hazumi}, {Holzapfel}, {Hori}, {Howard}, {Hyland}, {Inoue},
  {Jaehnig}, {Jaffe}, {Keating}, {Kermish}, {Keskitalo}, {Kisner}, {Le Jeune},
  {Lee}, {Leitch}, {Linder}, {Lungu}, {Matsuda}, {Matsumura}, {Meng}, {Miller},
  {Morii}, {Moyerman}, {Myers}, {Navaroli}, {Nishino}, {Paar}, {Peloton},
  {Poletti}, {Quealy}, {Rebeiz}, {Reichardt}, {Richards}, {Ross}, {Schanning},
  {Schenck}, {Sherwin}, {Shimizu}, {Shimmin}, {Shimon}, {Siritanasak},
  {Smecher}, {Spieler}, {Stebor}, {Steinbach}, {Stompor}, {Suzuki}, {Takakura},
  {Tomaru}, {Wilson}, {Yadav}, \& {Zahn}}]{PB2014BB}
{POLARBEAR Collaboration}, {Ade}, P.~A.~R., {Akiba}, Y., {et~al.} 2014, The
  Astrophysical Journal, 794, 171

\bibitem[{{Pryke} {et~al.}(2009){Pryke}, {Ade}, {Bock}, {Bowden}, {Brown},
  {Cahill}, {Castro}, {Church}, {Culverhouse}, {Friedman}, {Ganga}, {Gear},
  {Gupta}, {Hinderks}, {Kovac}, {Lange}, {Leitch}, {Melhuish}, {Memari},
  {Murphy}, {Orlando}, {Schwarz}, {O'Sullivan}, {Piccirillo}, {Rajguru},
  {Rusholme}, {Taylor}, {Thompson}, {Turner}, {Wu}, \&
  {Zemcov}}]{PrykeQUaD2009}
{Pryke}, C., {Ade}, P., {Bock}, J., {et~al.} 2009, \apj, 692, 1247

\bibitem[{{QUIET Collaboration} {et~al.}(2013){QUIET Collaboration},
  {Bischoff}, {Brizius}, {Buder}, {Chinone}, {Cleary}, {Dumoulin}, {Kusaka},
  {Monsalve}, {N{\ae}ss}, {Newburgh}, {Nixon}, {Reeves}, {Smith},
  {Vanderlinde}, {Wehus}, {Bogdan}, {Bustos}, {Church}, {Davis}, {Dickinson},
  {Eriksen}, {Gaier}, {Gundersen}, {Hasegawa}, {Hazumi}, {Holler},
  {Huffenberger}, {Imbriale}, {Ishidoshiro}, {Jones}, {Kangaslahti}, {Kapner},
  {Lawrence}, {Leitch}, {Limon}, {McMahon}, {Miller}, {Nagai}, {Nguyen},
  {Pearson}, {Piccirillo}, {Radford}, {Readhead}, {Richards}, {Samtleben},
  {Seiffert}, {Shepherd}, {Staggs}, {Tajima}, {Thompson}, {Williamson},
  {Winstein}, {Wollack}, \& {Zwart}}]{QUIETInstrument2013}
{QUIET Collaboration}, {Bischoff}, C., {Brizius}, A., {et~al.} 2013, \apj, 768,
  9

\bibitem[{{Reichborn-Kjennerud} {et~al.}(2010){Reichborn-Kjennerud},
  {Aboobaker}, {Ade}, {Aubin}, {Baccigalupi}, {Bao}, {Borrill}, {Cantalupo},
  {Chapman}, {Didier}, {Dobbs}, {Grain}, {Grainger}, {Hanany}, {Hillbrand},
  {Hubmayr}, {Jaffe}, {Johnson}, {Jones}, {Kisner}, {Klein}, {Korotkov},
  {Leach}, {Lee}, {Levinson}, {Limon}, {MacDermid}, {Matsumura}, {Meng},
  {Miller}, {Milligan}, {Pascale}, {Polsgrove}, {Ponthieu}, {Raach}, {Sagiv},
  {Smecher}, {Stivoli}, {Stompor}, {Tran}, {Tristram}, {Tucker}, {Vinokurov},
  {Yadav}, {Zaldarriaga}, \& {Zilic}}]{EBEx2010SPIE}
{Reichborn-Kjennerud}, B., {Aboobaker}, A.~M., {Ade}, P., {et~al.} 2010, in
  \procspie, Vol. 7741, Millimeter, Submillimeter, and Far-Infrared Detectors
  and Instrumentation for Astronomy V, 77411C

\bibitem[{{Revenu} {et~al.}(2000){Revenu}, {Kim}, {Ansari}, {Couchot},
  {Delabrouille}, \& {Kaplan}}]{Revenu2000}
{Revenu}, B., {Kim}, A., {Ansari}, R., {et~al.} 2000, \aaps, 142, 499

\bibitem[{{Shimon} {et~al.}(2008){Shimon}, {Keating}, {Ponthieu}, \&
  {Hivon}}]{Shimon2008}
{Shimon}, M., {Keating}, B., {Ponthieu}, N., \& {Hivon}, E. 2008, \prd, 77,
  083003

\bibitem[{{Smith}(2006)}]{KSmith2006}
{Smith}, K.~M. 2006, \prd, 74, 083002

\bibitem[{{Smoot} {et~al.}(1992){Smoot}, {Bennett}, {Kogut}, {Wright}, {Aymon},
  {Boggess}, {Cheng}, {de Amici}, {Gulkis}, {Hauser}, {Hinshaw}, {Jackson},
  {Janssen}, {Kaita}, {Kelsall}, {Keegstra}, {Lineweaver}, {Loewenstein},
  {Lubin}, {Mather}, {Meyer}, {Moseley}, {Murdock}, {Rokke}, {Silverberg},
  {Tenorio}, {Weiss}, \& {Wilkinson}}]{Smoot1992COBE}
{Smoot}, G.~F., {Bennett}, C.~L., {Kogut}, A., {et~al.} 1992, \apjl, 396, L1

\bibitem[{{Spinelli} {et~al.}(2011){Spinelli}, {Fabbian}, {Tartari}, {Zannoni},
  \& {Gervasi}}]{Spinelli2011}
{Spinelli}, S., {Fabbian}, G., {Tartari}, A., {Zannoni}, M., \& {Gervasi}, M.
  2011, \mnras, 414, 3272

\bibitem[{{Story} {et~al.}(2013){Story}, {Reichardt}, {Hou}, {Keisler}, {Aird},
  {Benson}, {Bleem}, {Carlstrom}, {Chang}, {Cho}, {Crawford}, {Crites}, {de
  Haan}, {Dobbs}, {Dudley}, {Follin}, {George}, {Halverson}, {Holder},
  {Holzapfel}, {Hoover}, {Hrubes}, {Joy}, {Knox}, {Lee}, {Leitch}, {Lueker},
  {Luong-Van}, {McMahon}, {Mehl}, {Meyer}, {Millea}, {Mohr}, {Montroy},
  {Padin}, {Plagge}, {Pryke}, {Ruhl}, {Sayre}, {Schaffer}, {Shaw}, {Shirokoff},
  {Spieler}, {Staniszewski}, {Stark}, {van Engelen}, {Vanderlinde}, {Vieira},
  {Williamson}, \& {Zahn}}]{StorySPT2013}
{Story}, K.~T., {Reichardt}, C.~L., {Hou}, Z., {et~al.} 2013, \apj, 779, 86

\bibitem[{{Tegmark}(1997)}]{Tegmark1997}
{Tegmark}, M. 1997, \prd, 56, 4514

\bibitem[{{Wollack} {et~al.}(1997){Wollack}, {Devlin}, {Jarosik},
  {Netterfield}, {Page}, \& {Wilkinson}}]{Wollack1997}
{Wollack}, E.~J., {Devlin}, M.~J., {Jarosik}, N., {et~al.} 1997, \apj, 476, 440

\bibitem[{{Wright} {et~al.}(1996){Wright}, {Hinshaw}, \&
  {Bennett}}]{Wright1996}
{Wright}, E.~L., {Hinshaw}, G., \& {Bennett}, C.~L. 1996, \apjl, 458, L53

\bibitem[{{Zaldarriaga} \& {Seljak}(1997)}]{Zaldarriaga1997}
{Zaldarriaga}, M., \& {Seljak}, U. 1997, \prd, 55, 1830

\bibitem[{{Zaldarriaga} \& {Seljak}(1998)}]{Zaldarriaga1998}
---. 1998, \prd, 58, 023003

\end{thebibliography}
\bibliographystyle{apj}

\end{document}